\pdfoutput=1
\documentclass[%
 reprint,
superscriptaddress,
 amsmath,amssymb,
 aps,
floatfix,
]{revtex4-1}

\usepackage{graphicx}% Include figure files
\usepackage{dcolumn}% Align table columns on decimal point
\usepackage{bm}% bold math
\usepackage{placeins}
\begin{document}

\preprint{APS/123-QED}

\title{Molecular and polymeric amorphous forms in dense SO\textsubscript{2}}

\author{Huichao Zhang}
\affiliation{Key Laboratory of Materials Physics, Institute of Solid State Physics, Chinese Academy of Sciences, Hefei, 230031, China}
\affiliation{University of Science and Technology of China, Hefei 230026,China}

\author{Ondrej T\'{o}th}
\affiliation{Department of Experimental Physics, Comenius University, 
Mlynsk\'{a} Dolina F2, 842 48 Bratislava, Slovakia}

\author{Xiao-Di Liu}
\email{xiaodi@issp.ac.cn}
\affiliation{Key Laboratory of Materials Physics, Institute of Solid State Physics, Chinese Academy of Sciences, Hefei, 230031, China}

\author{Roberto Bini} 
\affiliation{Department of Chemistry Univ Florence and European Laboratory for non Linear Spectroscopy (LENS), via N. Carrara 1, 50019 Sesto Fiorentino, Italy}

\author{Eugene Gregoryanz}
\affiliation{Key Laboratory of Materials Physics, Institute of Solid State Physics, Chinese Academy of Sciences, Hefei, 230031, China}
\affiliation{School of Physics and Astronomy and Centre for Science at Extreme Conditions, University of Edinburgh, Edinburgh EH9 3JZ, UK}
\affiliation{Center for High Pressure Science Technology Advanced Research, 1690 Cailun Road, Shanghai, 201203, China}

\author{Philip Dalladay-Simpson}
\affiliation{Center for High Pressure Science Technology Advanced Research, 1690 Cailun Road, Shanghai, 201203, China}

\author{Simone De Panfilis}
\affiliation{Centre for Life Nano Science, Istituto Italiano di Tecnologia, viale Regina Elena 291, 00161 Rome, Italy}

\author{Mario Santoro}
\email{santoro@lens.unifi.it}
\affiliation{Key Laboratory of Materials Physics, Institute of Solid State Physics, Chinese Academy of Sciences, Hefei, 230031, China}
\affiliation{Istituto Nazionale di Ottica (CNR-INO) and European Laboratory for non Linear Spectroscopy (LENS), via N. Carrara 1, 50019 Sesto Fiorentino, Italy}

\author{Federico Gorelli}
\email{gorelli@lens.unifi.it}
\affiliation{Key Laboratory of Materials Physics, Institute of Solid State Physics, Chinese Academy of Sciences, Hefei, 230031, China}
\affiliation{Istituto Nazionale di Ottica (CNR-INO) and European Laboratory for non Linear Spectroscopy (LENS), via N. Carrara 1, 50019 Sesto Fiorentino, Italy}

\author{Roman Marto\v{n}\'{a}k}
\email{martonak@fmph.uniba.sk}
\affiliation{Department of Experimental Physics, Comenius University, 
Mlynsk\'{a} Dolina F2, 842 48 Bratislava, Slovakia}

\date{\today}

\begin{abstract}
We report here a study of reversible pressure-induced structural transformation between two amorphous forms of SO\textsubscript{2}: molecular at pressures below 26 GPa and polymeric above this pressure, at temperatures of 77 - 300K.
The transformation was observed by Raman spectroscopy and x-ray diffraction in a diamond anvil cell. The same phenomenon was also observed in \textit{ab initio} molecular dynamics simulations where both forward and reverse transitions were detected, allowing to analyze in detail the atomic structure of both phases. The high-pressure polymeric amorphous form was found to consist mainly of disordered polymeric chains made of 3-coordinated sulfur atoms connected via oxygen atoms, and few residual intact molecules. The simulation results are in good agreement with experimental data. Our observations suggest a possible existence of molecular liquid - polymeric liquid transition in SO\textsubscript{2} and show a case example of polyamorphism in system consisting of simple molecules with multiple bonds.
\end{abstract}

\maketitle
Polyamorphism  is the counterpart of polymorphism observed in crystalline solids. It is characterized by the existence of two or more disordered forms, either amorphous or liquid, differing in local structural order, but preserving the original stoichiometry. This phenomenon is often also accompanied by differences in coordination and density \cite{Polyamorphism_Poole,Polyamorphism_McMillan}, (for recent reviews see Refs.~\cite{Machon2014, Anisimov2018}). Transformations between different amorphous forms can be driven by pressure and temperature. While in the case of crystalline polymorphs the structural transitions connect two equilibrium states and are often of first order and (at least in principle) sharp, in amorphous systems they connect two non-equilibrium states and are not necessarily of any given order, since structural disorder and lack of constraints imposed by lattice periodicity allow for a continuous evolution and a gradual transformation even between very different forms. The first example of such polyamorphic behavior was discovered by Mishima in 1984 in water ice \cite{Ice_Mishima_1984,Ice_Mishima_1985},
where compression of ice I\textsubscript{h} at 77 K induced a transformation to amorphous state. 
By applying specific compression/decompression/heating protocols, at least two different forms of amorphous water ice were found, called low-density amorphous (LDA) and high-density amorphous (HDA). The local structural order in the HDA and LDA forms differs by the presence of nonbonded water molecules in the first coordination shell and the two forms have also substantially different density.  The existence of two amorphous ices was suggested to be related to existence of liquid-liquid transition and second critical point of water \cite{Poole1992}. 
Similar phenomena were observed also in other systems such as Si \cite{SiMcMillan2005}, SiO\textsubscript{2} \cite{SiO2_Grimsditch_1984,SiO2Hemley1988,SilicateWilliams1988}, GeO\textsubscript{2} \cite{GeO2_Durben_1991}, where  polyamorphism is related to the change from tetrahedral to octahedral coordination at high pressure, or S \cite{Sanloup_aS} (for other examples and review see Refs.~\cite{Machon2014, Anisimov2018}).
Similar dramatic structural changes leading to amorphization have been observed upon compression of molecular crystals where multiple bonds are present. It is well known that pressure can destabilize multiple bonds in molecules and create extended polymeric networks with higher coordination than in the molecular phase. Due to associated strong kinetic effects creation of amorphous phases is often observed, especially when compression is performed at low temperature.
Amorphization of molecular crystals at high pressure have been observed in the famous examples of nitrogen \cite{N2Goncharov2000,N2Gregoryanz2001}, carbon dioxide \cite{CO2Santoro_2006,CarboniaMontoya_2008} and benzene \cite{BenzeneCiabiniJCP2002,BenzeneCiabiniPRL2002,BenzeneCiabiniNatMat2007}. The strong triple bond of the nitrogen molecule breaks under high pressures giving rise to a single-bonded network, while in carbon dioxide the double bond becomes unstable and carbon coordination increases to 3 and 4. In the case of benzene the aromatic ring opens and a network of hydrogenated carbons with single bonds is formed. The parent crystalline states of these amorphous materials have been discovered both in nitrogen \cite{N2Mailhiot1992,N2Eremets2004} and in carbon dioxide \cite{CO2Datchi2012,CO2Santoro2012} after high temperature annealing obtained by means of laser heating.
 
Here we present a hitherto unobserved example of polyamorphism related to a transition between molecular and polymeric amorphous forms of SO\textsubscript{2}. Sulfur dioxide is an important molecule in chemistry and has a significant role in industrial applications as well as in atmospheric and geological processes.
Unlike CO$_2$, the SO\textsubscript{2} molecule has structure that can be represented as two resonance structures with one single and one double bond \cite{InorgChem_House,SO2_molecule}.
Molecular crystals of SO\textsubscript{2} at pressures up to 32 GPa have been experimentally studied previously (Ref.~\cite{SO2Song2005} and references therein).
We have recently performed new experimental investigations of SO\textsubscript{2} at high pressures up to 60 GPa by means of Raman spectroscopy and X-ray diffraction (XRD). We have collected many spectra along isothermal compression runs at low temperature followed by decompression runs at room temperature. We support the experimental data with \textit{ab initio} simulations providing full access to atomistic details of the transformations under compression/decompression cycles. We have observed changes in agreement with literature at low pressures \cite{SO2Song2005} but also new changes at higher pressures. 
In particular upon compression at low temperature we have found an amorphization of the molecular crystal above about 15 GPa. This is followed above 26 GPa by a transformation to another amorphous state, a polymeric one, corresponding to an increase of the sulfur coordination from 2 to 3. Interestingly, the sequence of these transformations results to be quite reversible at room temperature even if with small hysteresis.
The same structural evolution is also observed in \textit{ab initio} molecular dynamics simulations which provide vibrational density of states (VDOS) agreeing very well with the main features of the Raman spectra for both phases.
Also, the static structure factor of both amorphous forms from simulation agrees very well with the measured one. The agreement with simulations provides a deep and  robust insight into the physical processes taking place with pressure in this simple molecular system.

\begin{figure}[!htb]
	%\centering
	\includegraphics[width=.95\linewidth]{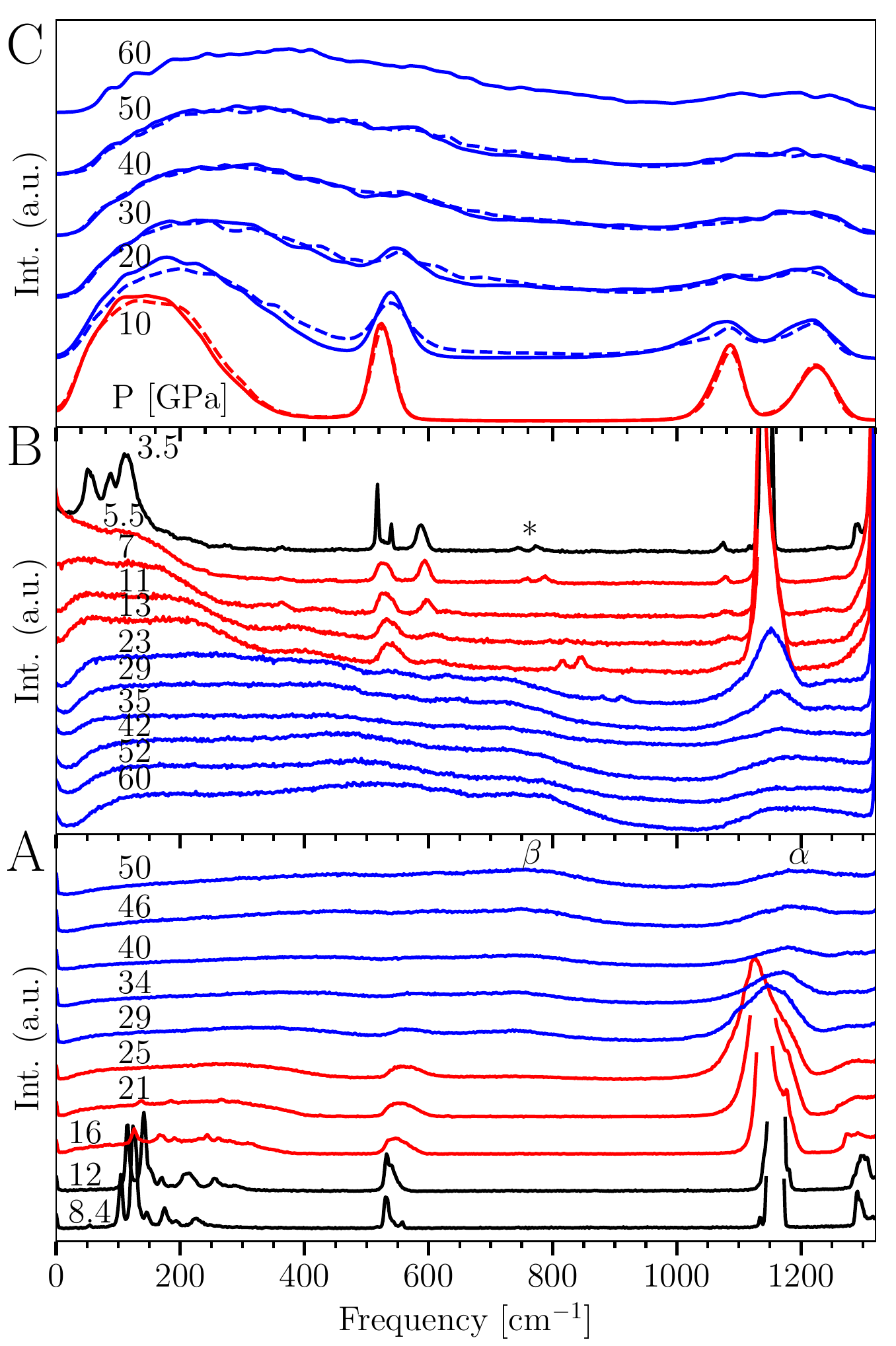}
	\caption{Vibrational spectra of solid SO\textsubscript{2}. Panels A and B: selected Raman spectra of an SO\textsubscript{2} sample measured upon increasing pressure at 77 K (A) and decreasing pressure at room temperature (B). During compression, the initially sharp molecular peaks of SO\textsubscript{2}, $\nu_1$, $\nu_2$, and $\nu_3$, broaden and become very weak while new broad and weak bands appear at different frequencies, indicating pressure-induced amorphization together with changes in the local structure. Upon decompression, the sharp molecular peaks of SO\textsubscript{2}, $\nu_1$, $\nu_2$, and $\nu_3$, are recovered while the new broad bands disappear at the same time, showing that amorphization and overall changes in the local structure are reversible. Panel C: evolution of vibrational density of states (VDOS) from ab initio MD simulations along compression (solid lines) and decompression (dashed lines). Color represents structural state of the system: black - molecular crystal, red - molecular amorphous and blue - polymeric amorphous. Star in panel B marks ruby peaks.
}
	\label{fig:raman}
\end{figure}

We start by presenting general features of Raman data observed upon compression/decompression. In Fig.~\ref{fig:raman}, we report selected Raman spectra measured upon increasing pressure up to 60 GPa, at 77 K (panel A), as well as upon decompression to ambient pressure at room temperature (panel B).
Similar results have been obtained for compressions at 210 K (Fig.~S1 Supp. Mat.). At a first glance, it immediately results that the sharp molecular peaks of SO\textsubscript{2}, i. e. $\nu_1$
1050-1220 cm\textsuperscript{-1}, $\nu_3$ 1240-1320 cm\textsuperscript{-1} and $\nu_2$ 520-600 cm\textsuperscript{-1}, observed at the lowest pressures broaden and become very weak upon increasing pressure.
In addition, new broad and weak bands appear at different frequencies, compatible with pressure-induced amorphization together with major changes in the local structure. On the other hand, the sharp molecular peaks of SO\textsubscript{2} are recovered upon decreasing pressure, while the new broad bands disappear, showing that these changes are indeed reversible. Specifically, reversibility strongly suggests that potential chemical decomposition into S and O\textsubscript{2} in our compressed samples is very likely to be ruled out, as decomposition would be rather an irreversible process. We then found a general agreement between our Raman spectra of solid molecular SO\textsubscript{2} under pressure and those previously measured at similar pressures \cite{SO2Song2005} in the common pressure range. On the other hand, dramatic spectral changes at low temperatures were not observed in the previous study \cite{SO2Song2005}, most likely because, in that study, pressure was limited to only 22 GPa, at low temperature.

If we now proceed with a closer look to the spectra, we can shed light on a number of remarkable details. We focus first on changes observed upon increasing pressure. Here we can first notice that the many sharp lattice peaks observed below 300-400 cm\textsuperscript{-1} weaken above 16 GPa till they completely disappear above 21 GPa, at 77 K (Fig.~\ref{fig:raman}~A), and above 10 GPa, at 210 K (Fig.S1 Supp. Mat.). Correspondingly, starting at around 16 GPa, at 77 K (10 GPa, at 210 K), a new broad band appears in the very same spectral region, while at higher frequencies we do not observe additional peaks other than the $\nu_1$, $\nu_2$, and $\nu_3$ molecular peaks of SO\textsubscript{2}, up to about 25 GPa. These peaks split into several Davydov components. The splitting and, even more importantly here, the linewidth of the components severely increase upon growing pressure while the intensity decreases substantially. All these facts straightforwardly suggest that upon cold compression above 10-15 GPa (depending on temperature) crystalline molecular SO\textsubscript{2} undergoes a structural transformation into a likely amorphous form, still entirely molecular, i.~e. made of separate SO\textsubscript{2} units with S in twofold coordination by O. Amorphization could have been enhanced by the shear stress, which is in turn related to the deformation of the gasket hole. This is supported also by our DFT calculations (see section on simulation results).
On the other hand, more dramatic changes occur above 22-25 GPa. Indeed, when pressure is increased further, the $\nu_2$ and $\nu_3$ peaks progressively disappear and they are hardly visible above 30-34 GPa. Instead, a single peak for the $\nu_1$ band, which is always by far the strongest molecular band, is still visible at our maximum pressures of 50-60 GPa (see also Fig.~\ref{fig:raman}~B), although very weak, positioned at 1170-1200 cm\textsuperscript{-1}. This peak is seemingly merged with an altogether new weak and broad peak centered at around 1220-1230 cm\textsuperscript{-1}, which we call peak $ \alpha $. In addition to this peak $ \alpha $, an entirely new, very broad band appears at 600-1000 cm\textsuperscript{-1}, with a high frequency edge at around 900 cm\textsuperscript{-1}, and we call this band $ \beta $. 
Both bands $ \alpha $ and $ \beta $ rather clearly appear to be of non-molecular origin and we suggest that these bands signal that amorphous, molecular SO\textsubscript{2} undergoes a transformation into a new amorphous, non-molecular/extended form of the same substance. Comparison with the pressure-induced molecular to amorphous-non-molecular transformation in CO\textsubscript{2} \cite{CO2Santoro_2006,CO2Santoro2012,CarboniaMontoya_2008} can help to interpret the transformation observed here in SO\textsubscript{2} since it bears some similarity. Indeed, carbonia, the amorphous, non-molecular CO\textsubscript{2} has been shown to be made of a mixture of 3-fold and 4-fold coordinated C by oxygen in similar proportions. 3-fold coordinated C sites are uniquely identified by C=O stretching peaks in the Raman and the IR spectrum at around 1900-2000 cm\textsuperscript{-1}, at 50-60 GPa, which roughly corresponds to the average value, $\nu$(stretch CO\textsubscript{2}), between the symmetric and the antisymmetric stretching modes of molecular CO\textsubscript{2}. 
On the other hand, considering the full Raman and IR spectra, the whole set of single C-O
bond stretching modes and deformation modes ascribed to both 3-fold and 4-fold coordinated C sites forms a broad spectral distribution extending in the 500-1500 cm\textsuperscript{-1} frequency range, 
which is about 0.26-0.77 of the average $\nu$(stretch CO\textsubscript{2}) frequency described above. If we now follow the same line of arguments for SO\textsubscript{2}, we immediately see that the above mentioned peak $ \alpha $ is positioned at roughly the average frequency between the two molecular stretching modes and thereby we can easily assign this peak to S=O stretching modes for non-molecular/extended SO\textsubscript{2} with S in 3-fold coordination by O. In addition, in the frequency range given by 0.26-0.77 of the frequency of peak $ \alpha $, that is 320-940 cm\textsuperscript{-1}, we indeed observe what we called above band $ \beta $. We can then straightforwardly assign band $ \beta $ to single S-O bond stretching modes and deformation modes of a form of non-molecular/extended SO\textsubscript{2}. 
At variance with carbonia, this form is now mainly made of S sites in 3-fold coordination by O, while the 4-fold coordination is absent, according to our DFT simulations (see later). We also need to consider that the $\nu_1$ peak for molecular SO\textsubscript{2} is still visible at our highest pressures, although very weak. We can thereby infer that the overall non-molecular/extended SO\textsubscript{2} obtained at pressures of tens of GPa is actually made of a mixture of some 2-fold S sites, still somewhat molecular in nature, and definitely non-molecular 3-fold coordinated S sites which represent the vast majority. An alternative possibility compatible with experimental data would be that molecular parts of the sample with 2-fold coordinated S and non-molecular parts with S in higher coordination are phase separated on a macroscopic/mesoscopic scale.

\begin{figure}[!htb]
	%\centering
	\includegraphics[width=.95\linewidth]{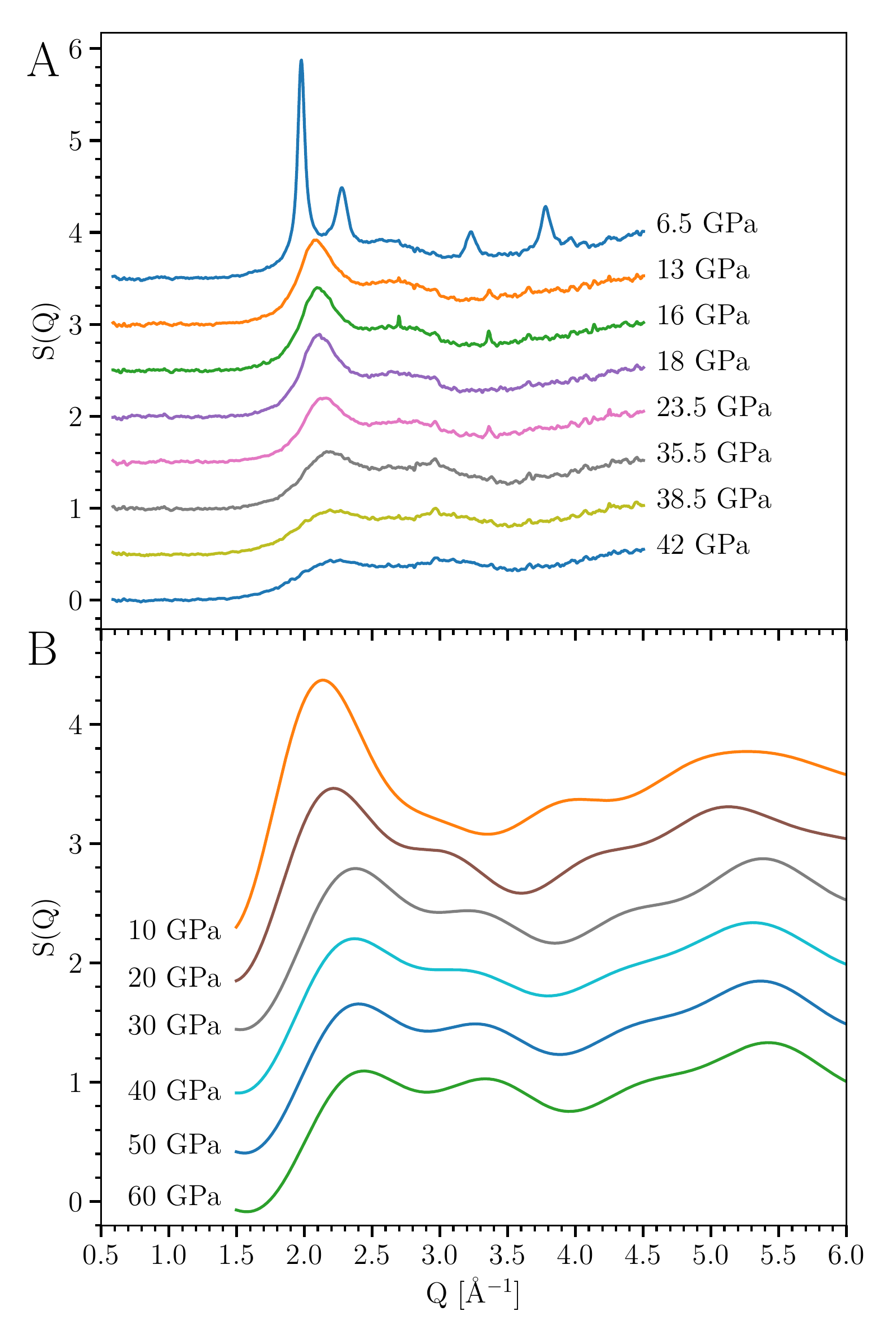}
	\caption{Static structure factor of solid SO\textsubscript{2} under pressure. Panel A: experimental $S(Q)$ measured along a room temperature decompression run, panel B: $S(Q)$ computed from simulations at 300 K during decompression. Region of $S(Q)$ beyond 4.5 \r{A}\textsuperscript{-1} in A is inaccessible because of limited angle in experiment, while the region below 1.5 \r{A}\textsuperscript{-1} in B is not reliable because of limited RDF range, resulting from the simulation supercell size of 14 \r{A}.}
	\label{fig:sq}
\end{figure}

Finally, Fig.~\ref{fig:raman}~B shows that the transformations found upon increasing pressure are indeed reverted upon decreasing pressure, although with small hysteresis, which is rather obvious since we deal here with kinetically limited structural changes. Specifically, we see that when pressure is lowered below 30-25 GPa, both peak $ \alpha $ and band $ \beta $ disappear rather suddenly, whereas the sharp molecular peaks $\nu_1$, $\nu_2$, and $\nu_3$ emerge suddenly as well. At the same time in the low frequency/lattice region below 350 cm\textsuperscript{-1} a diffuse, liquid-like band clearly develops, with no additional substantial changes down to about 5 GPa. These changes show that again an amorphous, yet molecular SO\textsubscript{2} form is obtained again at 25-5 GPa, similar to that found upon increasing pressure in a similar pressure range at low temperature. This form is then further transformed into crystalline, molecular SO\textsubscript{2} below 5 GPa, as indicated by the sharp lattice mode peaks observed below this pressure. An extra peak is also found below 20 GPa at around 600 cm\textsuperscript{-1}, similar to that observed in a previous study \cite{SO2Song2005} and assigned to possible molecular clusters. 
The reversible transformations of SO\textsubscript{2} under pressure to molecular and non-molecular amorphous forms to some extent parallel the similar case of CO\textsubscript{2} \cite{CO2Santoro_2006,CO2Santoro2012,CarboniaMontoya_2008}, at variance with the case of aromatic molecules where instead amorphous non-molecular forms obtained at high pressures are recovered at ambient conditions \cite{BenzeneCiabiniJCP2002,BenzeneCiabiniPRL2002,BenzeneCiabiniNatMat2007,FuraneCeppatelli,FuraneSantoro}.

The non-crystalline nature of SO\textsubscript{2} at high pressure has been assessed by XRD structural measurements at the Petra synchrotron facility. The evolution of the static structure factor upon decompression of the amorphous SO\textsubscript{2} sample previously obtained from a compression at low temperature is reported in Fig.~\ref{fig:sq}~A. The static structure factor has been obtained by the empty cell subtraction and taking into account the form factors of oxygen and sulfur and the Compton contribution from the sample following a procedure described elsewhere \cite{XRD_methodology}.
Between the patterns measured at 35.5 and 23.5 GPa in decompression, there is a clear change of the static structure factor corresponding to the transition from an extended amorphous to a molecular amorphous form. As a matter of fact at the highest pressures the two peaks at about 2 and 3 \r{A}\textsuperscript{-1} are of similar intensity, while on decompression the first peak significantly increases with respect to the second one which at the same time moves to the lower $Q$.

In order to obtain a better understanding of the microscopic processes on the atomistic level we performed \textit{ab initio} MD simulations following a pressure path similar to the experiment. We first performed a test in order to check whether applying shear stress to a perfect \textit{Aba} molecular crystal at low pressure might result in amorphous molecular structure as observed in experiment. 
We gradually induced shear strain by deforming the $\gamma$ angle of the supercell by up to 30$^{\circ}$ and indeed observed transformation into a disordered molecular form, confirming the experimentally observed amorphization.

\begin{figure}[!htb]
	\centering
%		\begin{figure}{\linewidth}
		\includegraphics[width=.95\linewidth]{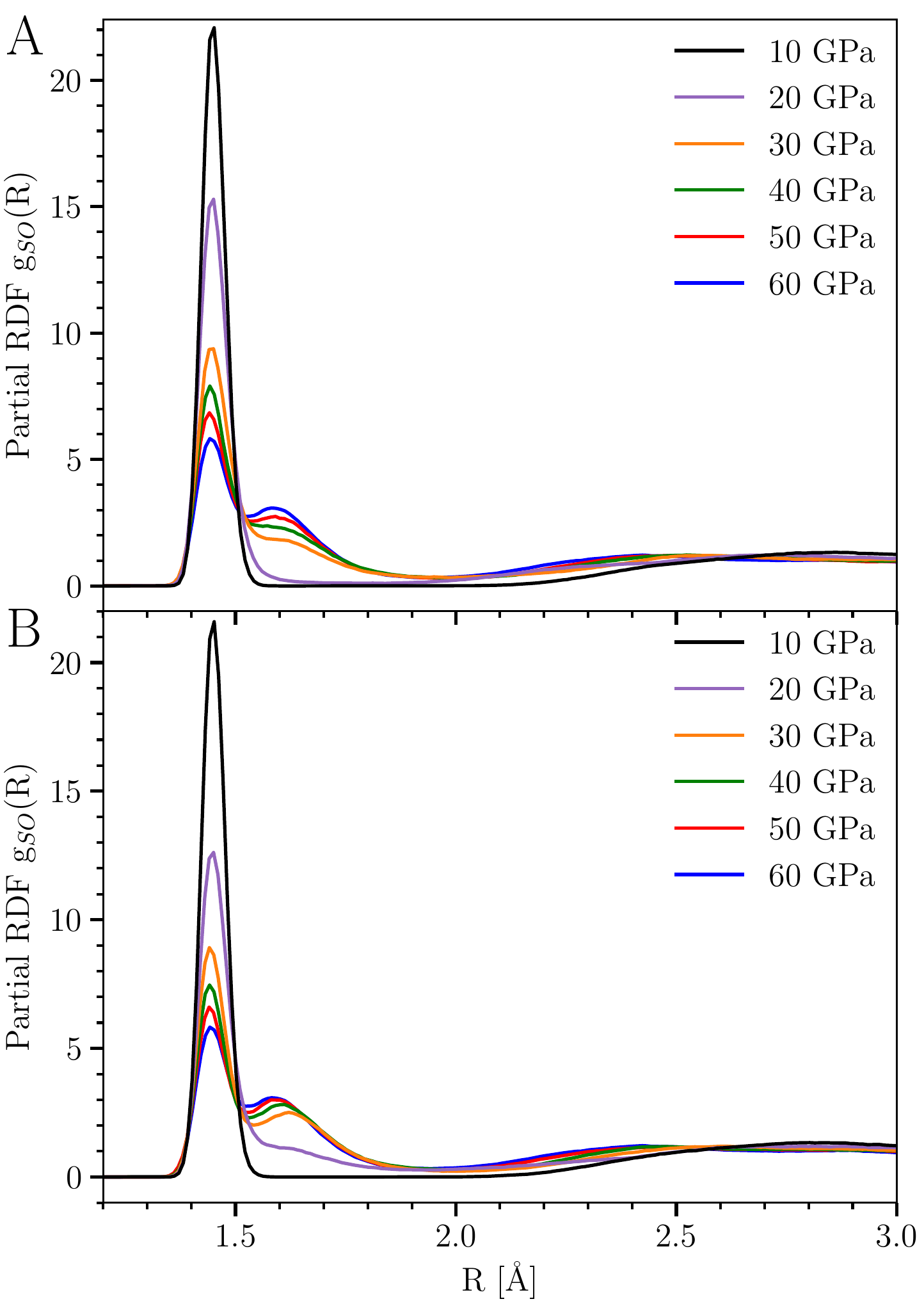}
%	\end{figure}
	\medskip
%	\begin{figure}{\linewidth}
		\includegraphics[width=.95\linewidth]{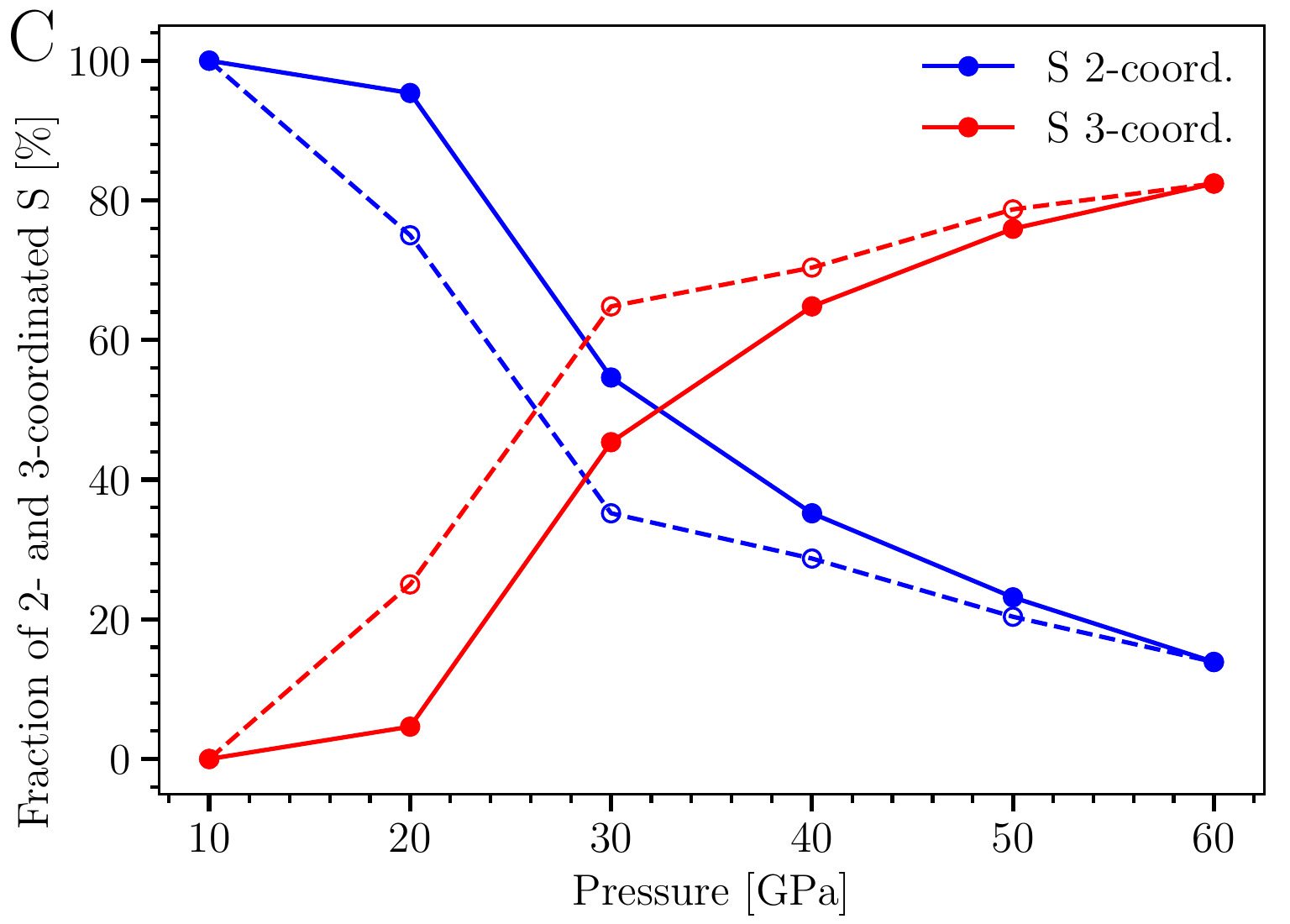}
%	\end{figure}
	\caption{Partial S-O radial distribution function (RDF) and concentration of sulfur coordination states. Panel A: RDF during compression, panel B: RDF during decompression and panel C: fraction of 2- and 3-coordinated S atoms during compression (solid line) and decompression (dashed line). Coordination number was determined within the cutoff of 1.92 \r{A}.}
	\label{fig:rdf}
\end{figure}

The full simulation protocol is shown in Fig.~S2 in Supp. Mat. In order to start the compression from well-defined amorphous molecular structure we melted a perfect \textit{Aba} molecular crystal \cite{SO2Post} in the $3\times3\times3$ supercell (108 SO\textsubscript{2} molecules, which equals to 324 atoms in unit cell) by heating it at $p=0$ to 600 K. By subsequent cooling down to 0 K we prepared the amorphous structure which served as starting point for further simulations.
We performed a gradual compression to 60 GPa and subsequent decompression to 10 GPa (in 10 GPa steps) at temperature of 300 K in order to accelerate the structural transformations (both compression and decompression). At this temperature we observed some diffusion of molecules in the molecular phase suggesting that the sample might possibly be in metastable liquid regime.

\begin{figure*}[!htb]%[tbhp]
    \centering
    \includegraphics[width=17.8cm,keepaspectratio]{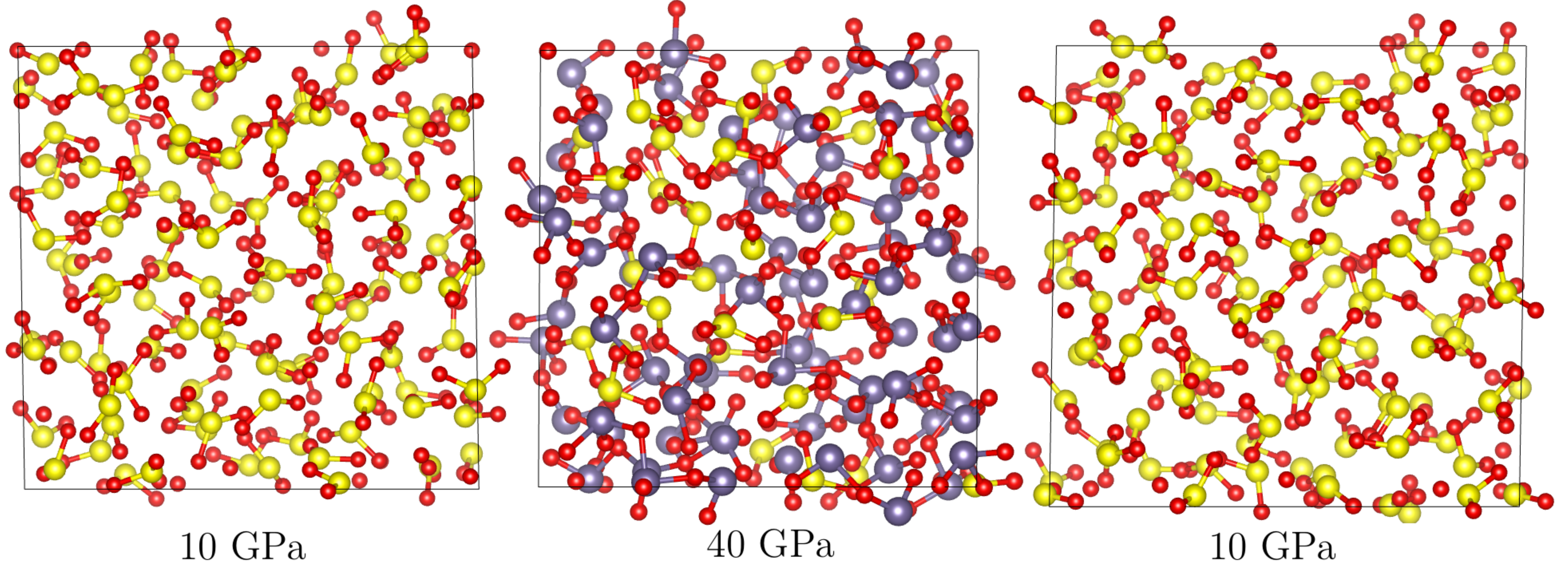}
    \caption{Snapshots of the simulated sample at different pressures. Left panel: the beginning of compression at 10 GPa where sample consists only of SO\textsubscript{2} molecules. Middle panel: structure during compression at 40 GPa with S atoms colored by coordination: 2-yellow (molecule) and 3-gray (polymeric chains). Right panel: structure after decompression to 10 GPa reverted from polymeric back to molecular. Simulation supercells are not to scale.}
    \label{fig:structure_evolution}
\end{figure*}

In order to provide a more direct description of the structure on the atomistic level, we first analyze the partial radial distribution functions (RDFs) obtained from simulations (Fig.~\ref{fig:rdf}).
These show an amorphous to amorphous transformation corresponding to the 2- to 3-fold S coordination change as well as the reverse transformation providing evidence for the reversibility of the structural change.
Upon compression of the initial molecular amorphous sample to 10 GPa at 300 K it still remains completely molecular as can be seen in Fig.~\ref{fig:rdf}~A. The peak at 1.44 \r{A} corresponding to the double bond is sharp and well separated from the next neighbor at 2.5-3.0 \r{A} and the coordination number of S atoms with respect to O atoms is 2. Upon compression to 20 GPa the RDF starts to change and at 30 GPa a substantial change can be seen. The first peak drops while a new peak at slightly longer distance of 1.6 \r{A} appears. This demonstrates that some of S=O double bonds are broken and replaced by single ones. At the same time 3-coordinated S-atoms appear with 2 single S-O bonds and one S=O double bond, forming polymeric chains (see snapshots shown in Fig.~\ref{fig:structure_evolution}). Upon increasing pressure the first peak progressively drops while the second one grows, resulting at 60 GPa in 82 \% of S atoms being 3-coordinated. Upon decompression (see Fig.~\ref{fig:rdf}~B) we observe a reverse evolution (see also the pressure dependence of the number of 2- and 3-coordinated S atoms in Fig.~\ref{fig:rdf}~C). At 10 GPa the polymeric chains disappear entirely (see Fig.~\ref{fig:structure_evolution}) and the system reverts back to molecular amorphous phase, in agreement with experiment. Simulations therefore very convincingly allow to interpret the experimental data as forward and backward transformation between molecular and polymeric amorphous forms. The dependence of the coordination number of S atoms on pressure (Fig.~\ref{fig:rdf}~C) exhibits some hysteresis which suggests that the transition might have a weakly first-order character. 
The pressure dependence of density upon compression and decompression, shown in Fig.~S3 (Supp. Mat.) shows a very small hysteresis and does not exhibit any particular features across the structural transformation.

The calculated static structure factor $ S(Q)$ shows important changes upon compression above 20 GPa (Fig.~S4 Supp. Mat.).
The intensity of the first diffraction peak at about 2.2 \r{A}\textsuperscript{-1} drops while a new peak appears in the region around 3 \r{A}\textsuperscript{-1} which grows upon increasing pressure. At 60 GPa the height of both peaks becomes similar, in agreement with experiment. 
All changes are reversible upon decompression with a small hysteresis and the calculated structure factor upon decompression agrees very well with the experimental one (Fig.~\ref{fig:sq}).
Simulations also allow us to decompose the total $S(Q)$ into contributions from atomic pairs (Figs.~S5 Supp. Mat.) The first peak at about 2 \r{A}\textsuperscript{-1} originates mainly from non-bonded S$\cdot\cdot\cdot$S pairs while the smaller peak at about 3 \r{A}\textsuperscript{-1} comes from O$\cdot\cdot\cdot$O and the broad one at 5 \r{A}\textsuperscript{-1} mainly from S-O pairs. However, it can be clearly seen that the drop of the first peak of the total $S(Q)$ at 2.2 \r{A}\textsuperscript{-1} is caused by the S-O contribution which in this region exhibits above 20 GPa a pronounced drop (Fig.~S5 Supp. Mat.). The observed evolution of $S(Q)$ therefore directly reflects the change of distance of S-O neighbors upon polymerization.

It is instructive to note here the analogy with selenium dioxide. According to the 9th rule of Prewitt and Downs \cite{Prewitt1998} light elements behave at high pressure like more heavy elements from the same group at lower pressures. Indeed, SeO\textsubscript{2} at ambient pressure forms polymeric chains ((O=Se)-O-)\textsubscript{n} (mineral downeyite), very similar to what we observe in SO\textsubscript{2} upon compression.

We now compare the vibrational spectra from simulations (Fig.~\ref{fig:raman}~C) with experimental Raman spectra (Fig.~\ref{fig:raman}).
Upon compression we observe qualitatively similar evolution to that found in the experiment (Fig.~\ref{fig:raman}~A). Above 20 GPa the distinct $\nu_2$ peak at 550 cm\textsuperscript{-1} progressively disappears while the background in the regions 400 - 500 and 600 - 900 cm\textsuperscript{-1} becomes gradually filled. At the same time the two molecular peaks, $\nu_1$ around 1100 cm\textsuperscript{-1} and $\nu_3$ above 1200 cm\textsuperscript{-1} gradually merge into a single broad peak around 1200 cm\textsuperscript{-1}, again in agreement with experiment (Fig.~\ref{fig:raman}~A). The evolution of the peaks can be understood from the projected VDOS allowing to decompose the total VDOS into contributions from structurally distinct S and O atoms (Fig.~S6 Supp. Mat.). The sulfur atoms are either in molecules or in polymeric chains. Oxygen atoms can be in molecules, or in polymeric chains, the latter ones being either in bridging position between two S atoms (S-O-S) or doubly bonded to S atoms (S=O). The evolution of the total VDOS clearly reflects the gradual conversion of molecules into polymeric chains. Upon decompression the reverse evolution is observed, again in agreement with experiment (Fig.~\ref{fig:raman}~B).

Finally, we analyzed from our DFT simulations the electronic properties of a-SO\textsubscript{2}. We found no metalization up to 60 GPa, with band gap of at least 0.6 eV in PBE approximation.

\subsection*{Conclusions}
We observed a reversible structural transition between molecular and polymeric amorphous forms of SO\textsubscript{2} at pressures around 26 GPa. The transition has rather small hysteresis pointing to the fact that the associated kinetic barriers are  low. The lower pressure of the transition between molecular and polymeric amorphous form, as well as the back transformation, is qualitatively facilitated by the molecular polarity which, supported by the high density attainable under pressure, drives the intermolecular interaction and lowers the activation energy of the transformation.
To our knowledge this kind of transition was not yet observed and provides a new example of structural transition between disordered non-equilibrium states of solid matter. 
Unlike in a-CO\textsubscript{2}, where polymeric a-carbonia contains 3- as well as 4-coordinated C atoms, here the molecular form converts into polymeric form with only 3-coordinated S atoms. 
The existence of molecular - polymeric transition in amorphous SO\textsubscript{2} suggests a possibility of an analogous transition in the liquid state, between molecular and polymeric liquids.
We note that in Ref.\cite{SO2Song2005}, the sample was in crystalline state at room temperature in the range of pressures from 2.5 to 32.7 GPa.
However, 
the position of the melting line of SO\textsubscript{2} under pressure is currently unknown and further experimental work is necessary to accurately map the solid and liquid regions of the phase diagram.

It is possible that such transitions between molecular and polymeric amorphous forms could be more generally present in molecular systems where multiple bonds are converted by pressure into polymeric chains or extended covalent solids. 
For example, in a recent computational study of amorphous polymeric nitrogen \cite{Melicherova2018} two dramatic structural changes were observed upon decompression. First, around 50 GPa the number of 3-coordinated atoms drops substantially and the system starts to create polymeric chains made of alternating single and double bonds. Such bonding is not realized in any stable crystalline phase of nitrogen. It shows that amorphous systems due to their intrinsic non-equilibrium nature may provide access to bonding patterns which are not observed in stable crystalline phases. Upon further decompression below 20 GPa the system starts to create molecules and transforms to molecular amorphous phase. Since in this case the kinetic barriers are arguably much larger, it is not possible to infer from simulations whether the evolution of structure and density is continuous or sudden changes occur. We believe that these interesting phenomena are worth further experimental investigation.

\subsection*{Matherials and methods}

\noindent\textbf{Experimental methodology.}
The SO\textsubscript{2} gas was loaded into the DAC by means of cryogenic loading: the gas was frozen between one diamond anvil and the gasket placed on the other diamond of a DAC which was opened by few mm and cooled to liquid nitrogen temperature inside a sealed glove box purged with nitrogen to avoid moisture condensation. 
We have performed Raman spectroscopy using a state of the art confocal Raman microscope with 15 and 2 micron of axial and transverse resolution respectively. The spectrometer is made of a Spectra Pro 750 mm monochromator, equipped with Pixis Princeton Instrument CCD detector. We have used Bragg Notch Filters to attenuate the laser light and spatial filtering of the collected light to obtain high quality spectra down to 7 cm\textsuperscript{-1} with minimal background from the diamond anvils and strong signal from the sample. The laser beam was expanded and cleaned by a Band Pass Filter. We have used a Laser Torus at 660 nm with 10 mW of power and Laser Ventus at 532 nm with 0.5 mW of power to check for the presence of eventual fluorescence bands in the spectrum. We generally used a 300 gr/mm grating as the spectral features were getting very broad and weak with pressure. The pressure was determined by the fluorescence of a small ruby placed in the sample or from the diamond stressed edge which we detect with high accuracy thanks to the excellent spatial resolution of the setup.
The XRD measurements were made at Petra (proposal ID: I-20181128) using monochromatic X-ray beam with 42.7 keV energy and the scattered X-rays were detected by a Perkin Elmer XRD1621 (2048x2048 pixels, 200x200um) detector. The diffraction patterns have been measured only along decompression of an amorphous sample obtained from a compression at low temperature while monitoring the changes with Raman spectroscopy. The excellent transverse spatial resolution allows to obtain clean diffraction patterns of the sample without the presence of spurious diffraction lines from the metallic gasket. The empty cell subtraction, which is of fundamental importance to obtain reliable measurements of the diffuse scattering from an amorphous or liquid sample in the DAC, have been in this case easily obtained by measuring the empty cell at the end of the decompression run when the SO\textsubscript{2} has completely back transformed to the gas state and escaped from the sample chamber.

\noindent\textbf{Simulations methodology.}
\textit{Ab initio} molecular dynamics simulations were performed by density functional theory (DFT) as implemented in VASP 5.3 and 5.4 codes \cite{VASP1,VASP2,VASP3}, employing projector augmented-wave pseudopotentials (with 6 valence electrons for both S and O atoms) and Perdew-Burke-Ernzerhof (PBE) \cite{PBE} parametrization of the GGA exchange-correlation functional. Compression, decompression and heating were performed by 6 ps variable cell NpT simulations with Langevin thermostat, energy cutoff 520 eV and $\Gamma$-point Brillouin zone sampling. We used 10.0 and 2.0 ps\textsuperscript{-1} friction coefficients for atomic and lattice degrees of freedom respectively, and 10000 $m_u$ as barostat fictitious mass. Data for velocity autocorrelation function were generated by equilibrating sample for 6 ps in NpT and then running 20 ps NVE simulation. Total and projected vibrational density of states (VDOS) were computed in standard way as Fourier transform of mass-weighted velocity autocorrelation function from MD trajectories at pressures from 10 to 60 GPa. Static structure factors $ S(Q) $ were calculated by performing Fourier transform of the RDFs from MD trajectories at several pressures along compression and decompression runs.

\begin{acknowledgments}
RM and OT were supported by the Slovak Research and
Development Agency under Contract No. APVV-15-0496. Calculations were
performed at the Computing Centre of the Slovak Academy of Sciences
using the supercomputing infrastructure acquired in ITMS Projects No.
26230120002 and No. 26210120002 (Slovak Infrastructure for
High-Performance Computing) supported by the Research and Development
Operational Programme funded by the ERDF.
XDL and HCZ were supported by National Natural Science Foundation of
China (Grant Nos.~11874361, 51672279, 11774354 and 51727806), Science
Challenge Project (No.~TZ2016001), and the CASHIPS Director’s Fund
(Grant No.~YZJJ2017705). FAG and MS were supported by CAS President’s
International Fellowship Initiative(PIFI), No.~2018VMA0053 and
No.~2019VMA0027 respectively.
\end{acknowledgments}

%\authorcontributions{R.M. designed the research, contributed to the analysis and discussion of results and wrote the paper. O.T. performed simulations, analyzed the data and wrote the paper. M.S. and F.A.G. designed the research, contributed to plan the experiments, to the data analysis and wrote the paper. P.D.S. performed the XRD measurement and contributed to finalizing the paper. S.D.P. performed XRD data analysis, contributed to the discussion of the results and to finalizing the paper. H.Z. performed the experiments and contributed to the data analysis. X.L. R.B. and E.G. contributed to the discussion of results and to finalizing the paper.}

%\authordeclaration{The authors declare no conflict of interest.}

\bibliographystyle{apsrev4-1}
\bibliography{citations.bib}

\begin{thebibliography}{37}%
\makeatletter
\providecommand \@ifxundefined [1]{%
 \@ifx{#1\undefined}
}%
\providecommand \@ifnum [1]{%
 \ifnum #1\expandafter \@firstoftwo
 \else \expandafter \@secondoftwo
 \fi
}%
\providecommand \@ifx [1]{%
 \ifx #1\expandafter \@firstoftwo
 \else \expandafter \@secondoftwo
 \fi
}%
\providecommand \natexlab [1]{#1}%
\providecommand \enquote  [1]{``#1''}%
\providecommand \bibnamefont  [1]{#1}%
\providecommand \bibfnamefont [1]{#1}%
\providecommand \citenamefont [1]{#1}%
\providecommand \href@noop [0]{\@secondoftwo}%
\providecommand \href [0]{\begingroup \@sanitize@url \@href}%
\providecommand \@href[1]{\@@startlink{#1}\@@href}%
\providecommand \@@href[1]{\endgroup#1\@@endlink}%
\providecommand \@sanitize@url [0]{\catcode `\\12\catcode `\$12\catcode
  `\&12\catcode `\#12\catcode `\^12\catcode `\_12\catcode `\%12\relax}%
\providecommand \@@startlink[1]{}%
\providecommand \@@endlink[0]{}%
\providecommand \url  [0]{\begingroup\@sanitize@url \@url }%
\providecommand \@url [1]{\endgroup\@href {#1}{\urlprefix }}%
\providecommand \urlprefix  [0]{URL }%
\providecommand \Eprint [0]{\href }%
\providecommand \doibase [0]{http://dx.doi.org/}%
\providecommand \selectlanguage [0]{\@gobble}%
\providecommand \bibinfo  [0]{\@secondoftwo}%
\providecommand \bibfield  [0]{\@secondoftwo}%
\providecommand \translation [1]{[#1]}%
\providecommand \BibitemOpen [0]{}%
\providecommand \bibitemStop [0]{}%
\providecommand \bibitemNoStop [0]{.\EOS\space}%
\providecommand \EOS [0]{\spacefactor3000\relax}%
\providecommand \BibitemShut  [1]{\csname bibitem#1\endcsname}%
\let\auto@bib@innerbib\@empty
%</preamble>
\bibitem [{\citenamefont {Poole}\ \emph {et~al.}(1997)\citenamefont {Poole},
  \citenamefont {Grande}, \citenamefont {Angell},\ and\ \citenamefont
  {McMillan}}]{Polyamorphism_Poole}%
  \BibitemOpen
  \bibfield  {author} {\bibinfo {author} {\bibfnamefont {P.~H.}\ \bibnamefont
  {Poole}}, \bibinfo {author} {\bibfnamefont {T.~T.}\ \bibnamefont {Grande}},
  \bibinfo {author} {\bibfnamefont {C.~A.}\ \bibnamefont {Angell}}, \ and\
  \bibinfo {author} {\bibfnamefont {P.~F.}\ \bibnamefont {McMillan}},\ }\href
  {\doibase 10.1126/science.275.5298.322} {\bibfield  {journal} {\bibinfo
  {journal} {Science}\ }\textbf {\bibinfo {volume} {275}},\ \bibinfo {pages}
  {322} (\bibinfo {year} {1997})},\ \Eprint
  {http://arxiv.org/abs/https://science.sciencemag.org/content}
  {https://science.sciencemag.org/content} \BibitemShut {NoStop}%
\bibitem [{\citenamefont {McMillan}(2004)}]{Polyamorphism_McMillan}%
  \BibitemOpen
  \bibfield  {author} {\bibinfo {author} {\bibfnamefont {P.~F.}\ \bibnamefont
  {McMillan}},\ }\href {\doibase 10.1039/B401308P} {\bibfield  {journal}
  {\bibinfo  {journal} {J. Mater. Chem.}\ }\textbf {\bibinfo {volume} {14}},\
  \bibinfo {pages} {1506} (\bibinfo {year} {2004})}\BibitemShut {NoStop}%
\bibitem [{\citenamefont {Machon}\ \emph {et~al.}(2014)\citenamefont {Machon},
  \citenamefont {Meersman}, \citenamefont {Wilding}, \citenamefont {Wilson},\
  and\ \citenamefont {McMillan}}]{Machon2014}%
  \BibitemOpen
  \bibfield  {author} {\bibinfo {author} {\bibfnamefont {D.}~\bibnamefont
  {Machon}}, \bibinfo {author} {\bibfnamefont {F.}~\bibnamefont {Meersman}},
  \bibinfo {author} {\bibfnamefont {M.}~\bibnamefont {Wilding}}, \bibinfo
  {author} {\bibfnamefont {M.}~\bibnamefont {Wilson}}, \ and\ \bibinfo {author}
  {\bibfnamefont {P.}~\bibnamefont {McMillan}},\ }\href {\doibase
  https://doi.org/10.1016/j.pmatsci.2013.12.002} {\bibfield  {journal}
  {\bibinfo  {journal} {Progress in Materials Science}\ }\textbf {\bibinfo
  {volume} {61}},\ \bibinfo {pages} {216 } (\bibinfo {year}
  {2014})}\BibitemShut {NoStop}%
\bibitem [{\citenamefont {Anisimov}\ \emph {et~al.}(2018)\citenamefont
  {Anisimov}, \citenamefont {Du\ifmmode~\check{s}\else \v{s}\fi{}ka},
  \citenamefont {Caupin}, \citenamefont {Amrhein}, \citenamefont {Rosenbaum},\
  and\ \citenamefont {Sadus}}]{Anisimov2018}%
  \BibitemOpen
  \bibfield  {author} {\bibinfo {author} {\bibfnamefont {M.~A.}\ \bibnamefont
  {Anisimov}}, \bibinfo {author} {\bibfnamefont {M.}~\bibnamefont
  {Du\ifmmode~\check{s}\else \v{s}\fi{}ka}}, \bibinfo {author} {\bibfnamefont
  {F.}~\bibnamefont {Caupin}}, \bibinfo {author} {\bibfnamefont {L.~E.}\
  \bibnamefont {Amrhein}}, \bibinfo {author} {\bibfnamefont {A.}~\bibnamefont
  {Rosenbaum}}, \ and\ \bibinfo {author} {\bibfnamefont {R.~J.}\ \bibnamefont
  {Sadus}},\ }\href {\doibase 10.1103/PhysRevX.8.011004} {\bibfield  {journal}
  {\bibinfo  {journal} {Phys. Rev. X}\ }\textbf {\bibinfo {volume} {8}},\
  \bibinfo {pages} {011004} (\bibinfo {year} {2018})}\BibitemShut {NoStop}%
\bibitem [{\citenamefont {Mishima}\ \emph {et~al.}(1984)\citenamefont
  {Mishima}, \citenamefont {Calvert},\ and\ \citenamefont
  {Whalley}}]{Ice_Mishima_1984}%
  \BibitemOpen
  \bibfield  {author} {\bibinfo {author} {\bibfnamefont {O.}~\bibnamefont
  {Mishima}}, \bibinfo {author} {\bibfnamefont {L.~D.}\ \bibnamefont
  {Calvert}}, \ and\ \bibinfo {author} {\bibfnamefont {E.}~\bibnamefont
  {Whalley}},\ }\href {\doibase 10.1038/310393a0} {\bibfield  {journal}
  {\bibinfo  {journal} {Nature}\ }\textbf {\bibinfo {volume} {310}},\ \bibinfo
  {pages} {393} (\bibinfo {year} {1984})}\BibitemShut {NoStop}%
\bibitem [{\citenamefont {Mishima}\ \emph {et~al.}(1985)\citenamefont
  {Mishima}, \citenamefont {Calvert},\ and\ \citenamefont
  {Whalley}}]{Ice_Mishima_1985}%
  \BibitemOpen
  \bibfield  {author} {\bibinfo {author} {\bibfnamefont {O.}~\bibnamefont
  {Mishima}}, \bibinfo {author} {\bibfnamefont {L.~D.}\ \bibnamefont
  {Calvert}}, \ and\ \bibinfo {author} {\bibfnamefont {E.}~\bibnamefont
  {Whalley}},\ }\href {\doibase 10.1038/314076a0} {\bibfield  {journal}
  {\bibinfo  {journal} {Nature}\ }\textbf {\bibinfo {volume} {314}},\ \bibinfo
  {pages} {76} (\bibinfo {year} {1985})}\BibitemShut {NoStop}%
\bibitem [{\citenamefont {Poole}\ \emph {et~al.}(1992)\citenamefont {Poole},
  \citenamefont {Sciortino}, \citenamefont {Essmann},\ and\ \citenamefont
  {Stanley}}]{Poole1992}%
  \BibitemOpen
  \bibfield  {author} {\bibinfo {author} {\bibfnamefont {P.~H.}\ \bibnamefont
  {Poole}}, \bibinfo {author} {\bibfnamefont {F.}~\bibnamefont {Sciortino}},
  \bibinfo {author} {\bibfnamefont {U.}~\bibnamefont {Essmann}}, \ and\
  \bibinfo {author} {\bibfnamefont {H.~E.}\ \bibnamefont {Stanley}},\ }\href
  {\doibase 10.1038/360324a0} {\bibfield  {journal} {\bibinfo  {journal}
  {Nature}\ }\textbf {\bibinfo {volume} {360}},\ \bibinfo {pages} {324}
  (\bibinfo {year} {1992})}\BibitemShut {NoStop}%
\bibitem [{\citenamefont {McMillan}\ \emph {et~al.}(2005)\citenamefont
  {McMillan}, \citenamefont {Wilson}, \citenamefont {Daisenberger},\ and\
  \citenamefont {Machon}}]{SiMcMillan2005}%
  \BibitemOpen
  \bibfield  {author} {\bibinfo {author} {\bibfnamefont {P.~F.}\ \bibnamefont
  {McMillan}}, \bibinfo {author} {\bibfnamefont {M.}~\bibnamefont {Wilson}},
  \bibinfo {author} {\bibfnamefont {D.}~\bibnamefont {Daisenberger}}, \ and\
  \bibinfo {author} {\bibfnamefont {D.}~\bibnamefont {Machon}},\ }\href
  {\doibase 10.1038/nmat1458} {\bibfield  {journal} {\bibinfo  {journal}
  {Nature}\ }\textbf {\bibinfo {volume} {4}},\ \bibinfo {pages} {680} (\bibinfo
  {year} {2005})}\BibitemShut {NoStop}%
\bibitem [{\citenamefont {Grimsditch}(1984)}]{SiO2_Grimsditch_1984}%
  \BibitemOpen
  \bibfield  {author} {\bibinfo {author} {\bibfnamefont {M.}~\bibnamefont
  {Grimsditch}},\ }\href {\doibase 10.1103/PhysRevLett.52.2379} {\bibfield
  {journal} {\bibinfo  {journal} {Physical Review Letters}\ }\textbf {\bibinfo
  {volume} {52}},\ \bibinfo {pages} {2379} (\bibinfo {year}
  {1984})}\BibitemShut {NoStop}%
\bibitem [{\citenamefont {Hemley}\ \emph {et~al.}(1988)\citenamefont {Hemley},
  \citenamefont {Jephcoat}, \citenamefont {Mao}, \citenamefont {Ming},\ and\
  \citenamefont {Manghnani}}]{SiO2Hemley1988}%
  \BibitemOpen
  \bibfield  {author} {\bibinfo {author} {\bibfnamefont {R.~J.}\ \bibnamefont
  {Hemley}}, \bibinfo {author} {\bibfnamefont {A.~P.}\ \bibnamefont
  {Jephcoat}}, \bibinfo {author} {\bibfnamefont {H.~K.}\ \bibnamefont {Mao}},
  \bibinfo {author} {\bibfnamefont {L.~C.}\ \bibnamefont {Ming}}, \ and\
  \bibinfo {author} {\bibfnamefont {M.~H.}\ \bibnamefont {Manghnani}},\ }\href
  {\doibase 10.1038/334052a0} {\bibfield  {journal} {\bibinfo  {journal}
  {Nature}\ }\textbf {\bibinfo {volume} {334}},\ \bibinfo {pages} {52}
  (\bibinfo {year} {1988})}\BibitemShut {NoStop}%
\bibitem [{\citenamefont {Williams}\ and\ \citenamefont
  {Jeanloz}(1988)}]{SilicateWilliams1988}%
  \BibitemOpen
  \bibfield  {author} {\bibinfo {author} {\bibfnamefont {Q.}~\bibnamefont
  {Williams}}\ and\ \bibinfo {author} {\bibfnamefont {R.}~\bibnamefont
  {Jeanloz}},\ }\href {\doibase 10.1126/science.239.4842.902} {\bibfield
  {journal} {\bibinfo  {journal} {Science}\ }\textbf {\bibinfo {volume}
  {239}},\ \bibinfo {pages} {902} (\bibinfo {year} {1988})},\ \Eprint
  {http://arxiv.org/abs/https://science.sciencemag.org/content/239/4842/902.full.pdf}
  {https://science.sciencemag.org/content/239/4842/902.full.pdf} \BibitemShut
  {NoStop}%
\bibitem [{\citenamefont {Durben}\ and\ \citenamefont
  {Wolf}(1991)}]{GeO2_Durben_1991}%
  \BibitemOpen
  \bibfield  {author} {\bibinfo {author} {\bibfnamefont {D.~J.}\ \bibnamefont
  {Durben}}\ and\ \bibinfo {author} {\bibfnamefont {G.~H.}\ \bibnamefont
  {Wolf}},\ }\href {\doibase 10.1103/PhysRevB.43.2355} {\bibfield  {journal}
  {\bibinfo  {journal} {Phys. Rev. B}\ }\textbf {\bibinfo {volume} {43}},\
  \bibinfo {pages} {2355} (\bibinfo {year} {1991})}\BibitemShut {NoStop}%
\bibitem [{\citenamefont {Sanloup}\ \emph {et~al.}(2008)\citenamefont
  {Sanloup}, \citenamefont {Gregoryanz}, \citenamefont {Degtyareva},\ and\
  \citenamefont {Hanfland}}]{Sanloup_aS}%
  \BibitemOpen
  \bibfield  {author} {\bibinfo {author} {\bibfnamefont {C.}~\bibnamefont
  {Sanloup}}, \bibinfo {author} {\bibfnamefont {E.}~\bibnamefont {Gregoryanz}},
  \bibinfo {author} {\bibfnamefont {O.}~\bibnamefont {Degtyareva}}, \ and\
  \bibinfo {author} {\bibfnamefont {M.}~\bibnamefont {Hanfland}},\ }\href
  {\doibase 10.1103/PhysRevLett.100.075701} {\bibfield  {journal} {\bibinfo
  {journal} {Phys. Rev. Lett.}\ }\textbf {\bibinfo {volume} {100}},\ \bibinfo
  {pages} {075701} (\bibinfo {year} {2008})}\BibitemShut {NoStop}%
\bibitem [{\citenamefont {Goncharov}\ \emph {et~al.}(2000)\citenamefont
  {Goncharov}, \citenamefont {Gregoryanz}, \citenamefont {Mao}, \citenamefont
  {Liu},\ and\ \citenamefont {Hemley}}]{N2Goncharov2000}%
  \BibitemOpen
  \bibfield  {author} {\bibinfo {author} {\bibfnamefont {A.~F.}\ \bibnamefont
  {Goncharov}}, \bibinfo {author} {\bibfnamefont {E.}~\bibnamefont
  {Gregoryanz}}, \bibinfo {author} {\bibfnamefont {H.}~\bibnamefont {Mao}},
  \bibinfo {author} {\bibfnamefont {Z.}~\bibnamefont {Liu}}, \ and\ \bibinfo
  {author} {\bibfnamefont {R.}~\bibnamefont {Hemley}},\ }\href {\doibase
  10.1103/PhysRevLett.85.1262} {\bibfield  {journal} {\bibinfo  {journal}
  {Physical review letters}\ }\textbf {\bibinfo {volume} {85}},\ \bibinfo
  {pages} {1262} (\bibinfo {year} {2000})}\BibitemShut {NoStop}%
\bibitem [{\citenamefont {Gregoryanz}\ \emph {et~al.}(2001)\citenamefont
  {Gregoryanz}, \citenamefont {Goncharov}, \citenamefont {Hemley},\ and\
  \citenamefont {Mao}}]{N2Gregoryanz2001}%
  \BibitemOpen
  \bibfield  {author} {\bibinfo {author} {\bibfnamefont {E.}~\bibnamefont
  {Gregoryanz}}, \bibinfo {author} {\bibfnamefont {A.~F.}\ \bibnamefont
  {Goncharov}}, \bibinfo {author} {\bibfnamefont {R.}~\bibnamefont {Hemley}}, \
  and\ \bibinfo {author} {\bibfnamefont {H.}~\bibnamefont {Mao}},\ }\href
  {\doibase 10.1103/PhysRevB.64.052103} {\bibfield  {journal} {\bibinfo
  {journal} {Physical Review B}\ }\textbf {\bibinfo {volume} {64}} (\bibinfo
  {year} {2001}),\ 10.1103/PhysRevB.64.052103}\BibitemShut {NoStop}%
\bibitem [{\citenamefont {Santoro}\ \emph {et~al.}(2006)\citenamefont
  {Santoro}, \citenamefont {Gorelli}, \citenamefont {Bini}, \citenamefont
  {Ruocco}, \citenamefont {Scandolo},\ and\ \citenamefont
  {Crichton}}]{CO2Santoro_2006}%
  \BibitemOpen
  \bibfield  {author} {\bibinfo {author} {\bibfnamefont {M.}~\bibnamefont
  {Santoro}}, \bibinfo {author} {\bibfnamefont {F.~A.}\ \bibnamefont
  {Gorelli}}, \bibinfo {author} {\bibfnamefont {R.}~\bibnamefont {Bini}},
  \bibinfo {author} {\bibfnamefont {G.}~\bibnamefont {Ruocco}}, \bibinfo
  {author} {\bibfnamefont {S.}~\bibnamefont {Scandolo}}, \ and\ \bibinfo
  {author} {\bibfnamefont {W.~A.}\ \bibnamefont {Crichton}},\ }\href {\doibase
  10.1038/nature04879} {\bibfield  {journal} {\bibinfo  {journal} {Nature}\
  }\textbf {\bibinfo {volume} {441}},\ \bibinfo {pages} {857} (\bibinfo {year}
  {2006})}\BibitemShut {NoStop}%
\bibitem [{\citenamefont {Montoya}\ \emph {et~al.}(2008)\citenamefont
  {Montoya}, \citenamefont {Rousseau}, \citenamefont {Santoro}, \citenamefont
  {Gorelli},\ and\ \citenamefont {Scandolo}}]{CarboniaMontoya_2008}%
  \BibitemOpen
  \bibfield  {author} {\bibinfo {author} {\bibfnamefont {J.~A.}\ \bibnamefont
  {Montoya}}, \bibinfo {author} {\bibfnamefont {R.}~\bibnamefont {Rousseau}},
  \bibinfo {author} {\bibfnamefont {M.}~\bibnamefont {Santoro}}, \bibinfo
  {author} {\bibfnamefont {F.}~\bibnamefont {Gorelli}}, \ and\ \bibinfo
  {author} {\bibfnamefont {S.}~\bibnamefont {Scandolo}},\ }\href {\doibase
  10.1103/PhysRevLett.100.163002} {\bibfield  {journal} {\bibinfo  {journal}
  {Phys. Rev. Lett.}\ }\textbf {\bibinfo {volume} {100}},\ \bibinfo {pages}
  {163002} (\bibinfo {year} {2008})}\BibitemShut {NoStop}%
\bibitem [{\citenamefont {Ciabini}\ \emph
  {et~al.}(2002{\natexlab{a}})\citenamefont {Ciabini}, \citenamefont {Santoro},
  \citenamefont {Bini},\ and\ \citenamefont
  {Schettino}}]{BenzeneCiabiniJCP2002}%
  \BibitemOpen
  \bibfield  {author} {\bibinfo {author} {\bibfnamefont {L.}~\bibnamefont
  {Ciabini}}, \bibinfo {author} {\bibfnamefont {M.}~\bibnamefont {Santoro}},
  \bibinfo {author} {\bibfnamefont {R.}~\bibnamefont {Bini}}, \ and\ \bibinfo
  {author} {\bibfnamefont {V.}~\bibnamefont {Schettino}},\ }\href {\doibase
  10.1063/1.1435570} {\bibfield  {journal} {\bibinfo  {journal} {The Journal of
  Chemical Physics}\ }\textbf {\bibinfo {volume} {116}},\ \bibinfo {pages}
  {2928} (\bibinfo {year} {2002}{\natexlab{a}})},\ \Eprint
  {http://arxiv.org/abs/https://doi.org/10.1063/1.1435570}
  {https://doi.org/10.1063/1.1435570} \BibitemShut {NoStop}%
\bibitem [{\citenamefont {Ciabini}\ \emph
  {et~al.}(2002{\natexlab{b}})\citenamefont {Ciabini}, \citenamefont {Santoro},
  \citenamefont {Bini},\ and\ \citenamefont
  {Schettino}}]{BenzeneCiabiniPRL2002}%
  \BibitemOpen
  \bibfield  {author} {\bibinfo {author} {\bibfnamefont {L.}~\bibnamefont
  {Ciabini}}, \bibinfo {author} {\bibfnamefont {M.}~\bibnamefont {Santoro}},
  \bibinfo {author} {\bibfnamefont {R.}~\bibnamefont {Bini}}, \ and\ \bibinfo
  {author} {\bibfnamefont {V.}~\bibnamefont {Schettino}},\ }\href {\doibase
  10.1103/PhysRevLett.88.085505} {\bibfield  {journal} {\bibinfo  {journal}
  {Physical review letters}\ }\textbf {\bibinfo {volume} {88}},\ \bibinfo
  {pages} {085505} (\bibinfo {year} {2002}{\natexlab{b}})}\BibitemShut
  {NoStop}%
\bibitem [{\citenamefont {Ciabini}\ \emph {et~al.}(2007)\citenamefont
  {Ciabini}, \citenamefont {Santoro}, \citenamefont {Gorelli}, \citenamefont
  {Bini}, \citenamefont {Schettino},\ and\ \citenamefont
  {Raugei}}]{BenzeneCiabiniNatMat2007}%
  \BibitemOpen
  \bibfield  {author} {\bibinfo {author} {\bibfnamefont {L.}~\bibnamefont
  {Ciabini}}, \bibinfo {author} {\bibfnamefont {M.}~\bibnamefont {Santoro}},
  \bibinfo {author} {\bibfnamefont {F.~A.}\ \bibnamefont {Gorelli}}, \bibinfo
  {author} {\bibfnamefont {R.}~\bibnamefont {Bini}}, \bibinfo {author}
  {\bibfnamefont {V.}~\bibnamefont {Schettino}}, \ and\ \bibinfo {author}
  {\bibfnamefont {S.}~\bibnamefont {Raugei}},\ }\href {\doibase
  10.1038/nmat1803} {\bibfield  {journal} {\bibinfo  {journal} {Nature}\
  }\textbf {\bibinfo {volume} {6}},\ \bibinfo {pages} {39} (\bibinfo {year}
  {2007})}\BibitemShut {NoStop}%
\bibitem [{\citenamefont {Mailhiot}\ \emph {et~al.}(1992)\citenamefont
  {Mailhiot}, \citenamefont {Yang},\ and\ \citenamefont
  {McMahan}}]{N2Mailhiot1992}%
  \BibitemOpen
  \bibfield  {author} {\bibinfo {author} {\bibfnamefont {C.}~\bibnamefont
  {Mailhiot}}, \bibinfo {author} {\bibfnamefont {L.~H.}\ \bibnamefont {Yang}},
  \ and\ \bibinfo {author} {\bibfnamefont {A.~K.}\ \bibnamefont {McMahan}},\
  }\href {\doibase 10.1103/PhysRevB.46.14419} {\bibfield  {journal} {\bibinfo
  {journal} {Phys. Rev. B}\ }\textbf {\bibinfo {volume} {46}},\ \bibinfo
  {pages} {14419} (\bibinfo {year} {1992})}\BibitemShut {NoStop}%
\bibitem [{\citenamefont {Eremets}\ \emph {et~al.}(2004)\citenamefont
  {Eremets}, \citenamefont {Gavriliuk}, \citenamefont {Trojan}, \citenamefont
  {Dzivenko},\ and\ \citenamefont {Boehler}}]{N2Eremets2004}%
  \BibitemOpen
  \bibfield  {author} {\bibinfo {author} {\bibfnamefont {M.~I.}\ \bibnamefont
  {Eremets}}, \bibinfo {author} {\bibfnamefont {A.~G.}\ \bibnamefont
  {Gavriliuk}}, \bibinfo {author} {\bibfnamefont {I.~A.}\ \bibnamefont
  {Trojan}}, \bibinfo {author} {\bibfnamefont {D.~A.}\ \bibnamefont
  {Dzivenko}}, \ and\ \bibinfo {author} {\bibfnamefont {R.}~\bibnamefont
  {Boehler}},\ }\href {\doibase 10.1038/nmat1146} {\bibfield  {journal}
  {\bibinfo  {journal} {Nature}\ }\textbf {\bibinfo {volume} {3}},\ \bibinfo
  {pages} {558} (\bibinfo {year} {2004})}\BibitemShut {NoStop}%
\bibitem [{\citenamefont {Datchi}\ \emph {et~al.}(2012)\citenamefont {Datchi},
  \citenamefont {Mallick}, \citenamefont {Salamat},\ and\ \citenamefont
  {Ninet}}]{CO2Datchi2012}%
  \BibitemOpen
  \bibfield  {author} {\bibinfo {author} {\bibfnamefont {F.}~\bibnamefont
  {Datchi}}, \bibinfo {author} {\bibfnamefont {B.}~\bibnamefont {Mallick}},
  \bibinfo {author} {\bibfnamefont {A.}~\bibnamefont {Salamat}}, \ and\
  \bibinfo {author} {\bibfnamefont {S.}~\bibnamefont {Ninet}},\ }\href
  {\doibase 10.1103/PhysRevLett.108.125701} {\bibfield  {journal} {\bibinfo
  {journal} {Phys. Rev. Lett.}\ }\textbf {\bibinfo {volume} {108}},\ \bibinfo
  {pages} {125701} (\bibinfo {year} {2012})}\BibitemShut {NoStop}%
\bibitem [{\citenamefont {Santoro}\ \emph {et~al.}(2012)\citenamefont
  {Santoro}, \citenamefont {Gorelli}, \citenamefont {Bini}, \citenamefont
  {Haines}, \citenamefont {Cambon}, \citenamefont {Levelut}, \citenamefont
  {Montoya},\ and\ \citenamefont {Scandolo}}]{CO2Santoro2012}%
  \BibitemOpen
  \bibfield  {author} {\bibinfo {author} {\bibfnamefont {M.}~\bibnamefont
  {Santoro}}, \bibinfo {author} {\bibfnamefont {F.~A.}\ \bibnamefont
  {Gorelli}}, \bibinfo {author} {\bibfnamefont {R.}~\bibnamefont {Bini}},
  \bibinfo {author} {\bibfnamefont {J.}~\bibnamefont {Haines}}, \bibinfo
  {author} {\bibfnamefont {O.}~\bibnamefont {Cambon}}, \bibinfo {author}
  {\bibfnamefont {C.}~\bibnamefont {Levelut}}, \bibinfo {author} {\bibfnamefont
  {J.~A.}\ \bibnamefont {Montoya}}, \ and\ \bibinfo {author} {\bibfnamefont
  {S.}~\bibnamefont {Scandolo}},\ }\href {\doibase 10.1073/pnas.1118791109}
  {\bibfield  {journal} {\bibinfo  {journal} {Proceedings of the National
  Academy of Sciences}\ }\textbf {\bibinfo {volume} {109}},\ \bibinfo {pages}
  {5176} (\bibinfo {year} {2012})},\ \Eprint
  {http://arxiv.org/abs/https://www.pnas.org/content/109/14/5176.full.pdf}
  {https://www.pnas.org/content/109/14/5176.full.pdf} \BibitemShut {NoStop}%
\bibitem [{\citenamefont {House}(2008)}]{InorgChem_House}%
  \BibitemOpen
  \bibfield  {author} {\bibinfo {author} {\bibfnamefont {J.~E.}\ \bibnamefont
  {House}},\ }\href@noop {} {\emph {\bibinfo {title} {Inorganic Chemistry}}}\
  (\bibinfo  {publisher} {Academic Press},\ \bibinfo {year} {2008})\BibitemShut
  {NoStop}%
\bibitem [{\citenamefont {Takeshita}\ \emph {et~al.}(2015)\citenamefont
  {Takeshita}, \citenamefont {Lindquist},\ and\ \citenamefont
  {Dunning}}]{SO2_molecule}%
  \BibitemOpen
  \bibfield  {author} {\bibinfo {author} {\bibfnamefont {T.~Y.}\ \bibnamefont
  {Takeshita}}, \bibinfo {author} {\bibfnamefont {B.~A.}\ \bibnamefont
  {Lindquist}}, \ and\ \bibinfo {author} {\bibfnamefont {T.~H.}\ \bibnamefont
  {Dunning}},\ }\href {\doibase 10.1021/acs.jpca.5b00998} {\bibfield  {journal}
  {\bibinfo  {journal} {The Journal of Physical Chemistry A}\ }\textbf
  {\bibinfo {volume} {119}},\ \bibinfo {pages} {7683} (\bibinfo {year}
  {2015})},\ \bibinfo {note} {pMID: 26068052},\ \Eprint
  {http://arxiv.org/abs/https://doi.org/10.1021/acs.jpca.5b00998}
  {https://doi.org/10.1021/acs.jpca.5b00998} \BibitemShut {NoStop}%
\bibitem [{\citenamefont {Song}\ \emph {et~al.}(2005)\citenamefont {Song},
  \citenamefont {Liu}, \citenamefont {Mao}, \citenamefont {Hemley},\ and\
  \citenamefont {Herschbach}}]{SO2Song2005}%
  \BibitemOpen
  \bibfield  {author} {\bibinfo {author} {\bibfnamefont {Y.}~\bibnamefont
  {Song}}, \bibinfo {author} {\bibfnamefont {Z.}~\bibnamefont {Liu}}, \bibinfo
  {author} {\bibfnamefont {H.}~\bibnamefont {Mao}}, \bibinfo {author}
  {\bibfnamefont {R.~J.}\ \bibnamefont {Hemley}}, \ and\ \bibinfo {author}
  {\bibfnamefont {D.~R.}\ \bibnamefont {Herschbach}},\ }\href {\doibase
  10.1063/1.1883405} {\bibfield  {journal} {\bibinfo  {journal} {The Journal of
  Chemical Physics}\ }\textbf {\bibinfo {volume} {122}},\ \bibinfo {pages}
  {174511} (\bibinfo {year} {2005})},\ \Eprint
  {http://arxiv.org/abs/https://doi.org/10.1063/1.1883405}
  {https://doi.org/10.1063/1.1883405} \BibitemShut {NoStop}%
\bibitem [{\citenamefont {Ceppatelli}\ \emph {et~al.}(2003)\citenamefont
  {Ceppatelli}, \citenamefont {Santoro}, \citenamefont {Bini},\ and\
  \citenamefont {Schettino}}]{FuraneCeppatelli}%
  \BibitemOpen
  \bibfield  {author} {\bibinfo {author} {\bibfnamefont {M.}~\bibnamefont
  {Ceppatelli}}, \bibinfo {author} {\bibfnamefont {M.}~\bibnamefont {Santoro}},
  \bibinfo {author} {\bibfnamefont {R.}~\bibnamefont {Bini}}, \ and\ \bibinfo
  {author} {\bibfnamefont {V.}~\bibnamefont {Schettino}},\ }\href {\doibase
  10.1063/1.1527895} {\bibfield  {journal} {\bibinfo  {journal} {The Journal of
  Chemical Physics}\ }\textbf {\bibinfo {volume} {118}},\ \bibinfo {pages}
  {1499} (\bibinfo {year} {2003})},\ \Eprint
  {http://arxiv.org/abs/https://doi.org/10.1063/1.1527895}
  {https://doi.org/10.1063/1.1527895} \BibitemShut {NoStop}%
\bibitem [{\citenamefont {Santoro}\ \emph {et~al.}(2003)\citenamefont
  {Santoro}, \citenamefont {Ceppatelli}, \citenamefont {Bini},\ and\
  \citenamefont {Schettino}}]{FuraneSantoro}%
  \BibitemOpen
  \bibfield  {author} {\bibinfo {author} {\bibfnamefont {M.}~\bibnamefont
  {Santoro}}, \bibinfo {author} {\bibfnamefont {M.}~\bibnamefont {Ceppatelli}},
  \bibinfo {author} {\bibfnamefont {R.}~\bibnamefont {Bini}}, \ and\ \bibinfo
  {author} {\bibfnamefont {V.}~\bibnamefont {Schettino}},\ }\href {\doibase
  10.1063/1.1565997} {\bibfield  {journal} {\bibinfo  {journal} {The Journal of
  Chemical Physics}\ }\textbf {\bibinfo {volume} {118}},\ \bibinfo {pages}
  {8321} (\bibinfo {year} {2003})},\ \Eprint
  {http://arxiv.org/abs/https://doi.org/10.1063/1.1565997}
  {https://doi.org/10.1063/1.1565997} \BibitemShut {NoStop}%
\bibitem [{\citenamefont {Eggert}\ \emph {et~al.}(2002)\citenamefont {Eggert},
  \citenamefont {Weck}, \citenamefont {Loubeyre},\ and\ \citenamefont
  {Mezouar}}]{XRD_methodology}%
  \BibitemOpen
  \bibfield  {author} {\bibinfo {author} {\bibfnamefont {J.~H.}\ \bibnamefont
  {Eggert}}, \bibinfo {author} {\bibfnamefont {G.}~\bibnamefont {Weck}},
  \bibinfo {author} {\bibfnamefont {P.}~\bibnamefont {Loubeyre}}, \ and\
  \bibinfo {author} {\bibfnamefont {M.}~\bibnamefont {Mezouar}},\ }\href
  {\doibase 10.1103/PhysRevB.65.174105} {\bibfield  {journal} {\bibinfo
  {journal} {Phys. Rev. B}\ }\textbf {\bibinfo {volume} {65}},\ \bibinfo
  {pages} {174105} (\bibinfo {year} {2002})}\BibitemShut {NoStop}%
\bibitem [{\citenamefont {Post}\ \emph {et~al.}(1952)\citenamefont {Post},
  \citenamefont {Schwartz},\ and\ \citenamefont {Fankuchen}}]{SO2Post}%
  \BibitemOpen
  \bibfield  {author} {\bibinfo {author} {\bibfnamefont {B.}~\bibnamefont
  {Post}}, \bibinfo {author} {\bibfnamefont {R.~S.}\ \bibnamefont {Schwartz}},
  \ and\ \bibinfo {author} {\bibfnamefont {I.}~\bibnamefont {Fankuchen}},\
  }\href {\doibase 10.1107/S0365110X5200109X} {\bibfield  {journal} {\bibinfo
  {journal} {Acta Crystallographica}\ }\textbf {\bibinfo {volume} {5}},\
  \bibinfo {pages} {372} (\bibinfo {year} {1952})}\BibitemShut {NoStop}%
\bibitem [{\citenamefont {Prewitt}\ and\ \citenamefont
  {Down}(1998)}]{Prewitt1998}%
  \BibitemOpen
  \bibfield  {author} {\bibinfo {author} {\bibfnamefont {C.~T.}\ \bibnamefont
  {Prewitt}}\ and\ \bibinfo {author} {\bibfnamefont {R.~T.}\ \bibnamefont
  {Down}},\ }\href@noop {} {\bibfield  {journal} {\bibinfo  {journal} {Rev.
  Mineral.}\ }\textbf {\bibinfo {volume} {37}},\ \bibinfo {pages} {283}
  (\bibinfo {year} {1998})}\BibitemShut {NoStop}%
\bibitem [{\citenamefont {Melicherov\'a}\ \emph {et~al.}(2018)\citenamefont
  {Melicherov\'a}, \citenamefont {Kohul\'ak}, \citenamefont
  {Pla\ifmmode~\check{s}\else \v{s}\fi{}ienka},\ and\ \citenamefont
  {Marto\ifmmode~\check{n}\else \v{n}\fi{}\'ak}}]{Melicherova2018}%
  \BibitemOpen
  \bibfield  {author} {\bibinfo {author} {\bibfnamefont {D.}~\bibnamefont
  {Melicherov\'a}}, \bibinfo {author} {\bibfnamefont {O.}~\bibnamefont
  {Kohul\'ak}}, \bibinfo {author} {\bibfnamefont {D.~c.~v.}\ \bibnamefont
  {Pla\ifmmode~\check{s}\else \v{s}\fi{}ienka}}, \ and\ \bibinfo {author}
  {\bibfnamefont {R.}~\bibnamefont {Marto\ifmmode~\check{n}\else
  \v{n}\fi{}\'ak}},\ }\href {\doibase 10.1103/PhysRevMaterials.2.103601}
  {\bibfield  {journal} {\bibinfo  {journal} {Phys. Rev. Materials}\ }\textbf
  {\bibinfo {volume} {2}},\ \bibinfo {pages} {103601} (\bibinfo {year}
  {2018})}\BibitemShut {NoStop}%
\bibitem [{\citenamefont {Kresse}\ and\ \citenamefont {Hafner}(1993)}]{VASP1}%
  \BibitemOpen
  \bibfield  {author} {\bibinfo {author} {\bibfnamefont {G.}~\bibnamefont
  {Kresse}}\ and\ \bibinfo {author} {\bibfnamefont {G.~J.}\ \bibnamefont
  {Hafner}},\ }\href@noop {} {\bibfield  {journal} {\bibinfo  {journal} {Phys.
  Rev. B}\ }\textbf {\bibinfo {volume} {47}},\ \bibinfo {pages} {558} (\bibinfo
  {year} {1993})}\BibitemShut {NoStop}%
\bibitem [{\citenamefont {Kresse}\ and\ \citenamefont
  {Furthmüller}(1996)}]{VASP2}%
  \BibitemOpen
  \bibfield  {author} {\bibinfo {author} {\bibfnamefont {G.}~\bibnamefont
  {Kresse}}\ and\ \bibinfo {author} {\bibfnamefont {J.}~\bibnamefont
  {Furthmüller}},\ }\href {\doibase
  https://doi.org/10.1016/0927-0256(96)00008-0} {\bibfield  {journal} {\bibinfo
   {journal} {Computational Materials Science}\ }\textbf {\bibinfo {volume}
  {6}},\ \bibinfo {pages} {15 } (\bibinfo {year} {1996})}\BibitemShut {NoStop}%
\bibitem [{\citenamefont {Kresse}\ and\ \citenamefont
  {Furthm\"uller}(1996)}]{VASP3}%
  \BibitemOpen
  \bibfield  {author} {\bibinfo {author} {\bibfnamefont {G.}~\bibnamefont
  {Kresse}}\ and\ \bibinfo {author} {\bibfnamefont {J.}~\bibnamefont
  {Furthm\"uller}},\ }\href {\doibase 10.1103/PhysRevB.54.11169} {\bibfield
  {journal} {\bibinfo  {journal} {Phys. Rev. B}\ }\textbf {\bibinfo {volume}
  {54}},\ \bibinfo {pages} {11169} (\bibinfo {year} {1996})}\BibitemShut
  {NoStop}%
\bibitem [{\citenamefont {Perdew}\ \emph {et~al.}(1996)\citenamefont {Perdew},
  \citenamefont {Burke},\ and\ \citenamefont {Ernzerhof}}]{PBE}%
  \BibitemOpen
  \bibfield  {author} {\bibinfo {author} {\bibfnamefont {J.~P.}\ \bibnamefont
  {Perdew}}, \bibinfo {author} {\bibfnamefont {K.}~\bibnamefont {Burke}}, \
  and\ \bibinfo {author} {\bibfnamefont {M.}~\bibnamefont {Ernzerhof}},\ }\href
  {\doibase 10.1103/physrevlett.77.3865} {\bibfield  {journal} {\bibinfo
  {journal} {Phys. Rev. Lett.}\ }\textbf {\bibinfo {volume} {77}},\ \bibinfo
  {pages} {3865} (\bibinfo {year} {1996})}\BibitemShut {NoStop}%
\end{thebibliography}%

%\widetext
\onecolumngrid
\clearpage
\begin{center}
\textbf{\large Supplemental Material: Molecular and polymeric amorphous forms in dense SO\textsubscript{2}}
\end{center}

%reset counting for Supp. Mat.
\setcounter{equation}{0}
\setcounter{figure}{0}
\setcounter{table}{0}
\setcounter{page}{1}
\makeatletter
\renewcommand{\theequation}{S\arabic{equation}}
\renewcommand{\thefigure}{S\arabic{figure}}
\renewcommand{\bibnumfmt}[1]{[S#1]}
\renewcommand{\citenumfont}[1]{S#1}

\vspace{3cm}
\begin{figure*}[tbhp]
\centering
\includegraphics[width=.95\linewidth]{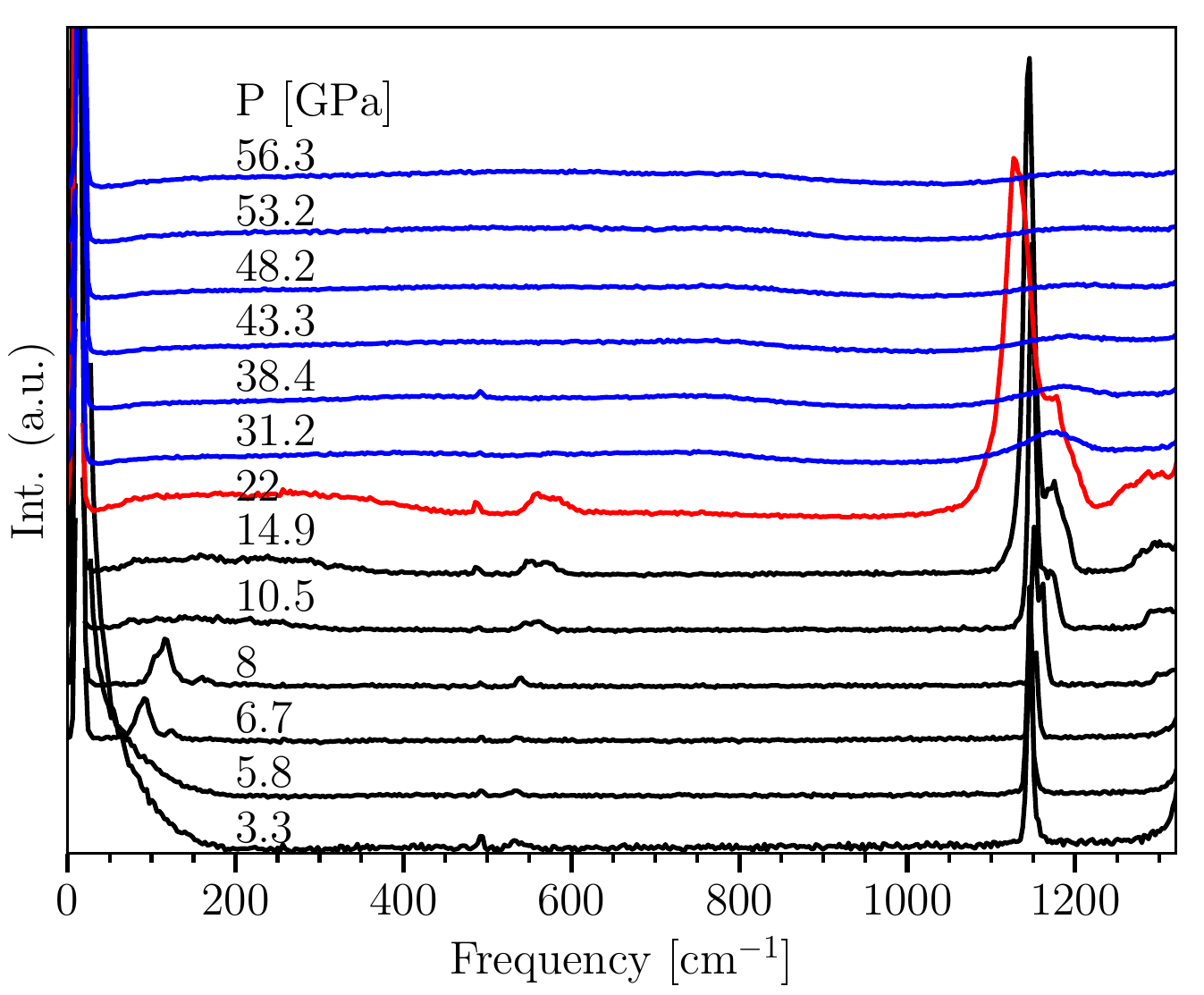}
\caption{Selected Raman spectra of an SO\textsubscript{2} sample upon increasing pressure at 210 K.}
\label{fig:raman_210}
\end{figure*}

\begin{figure*}%[tbhp]
\centering
\includegraphics[width=.95\linewidth]{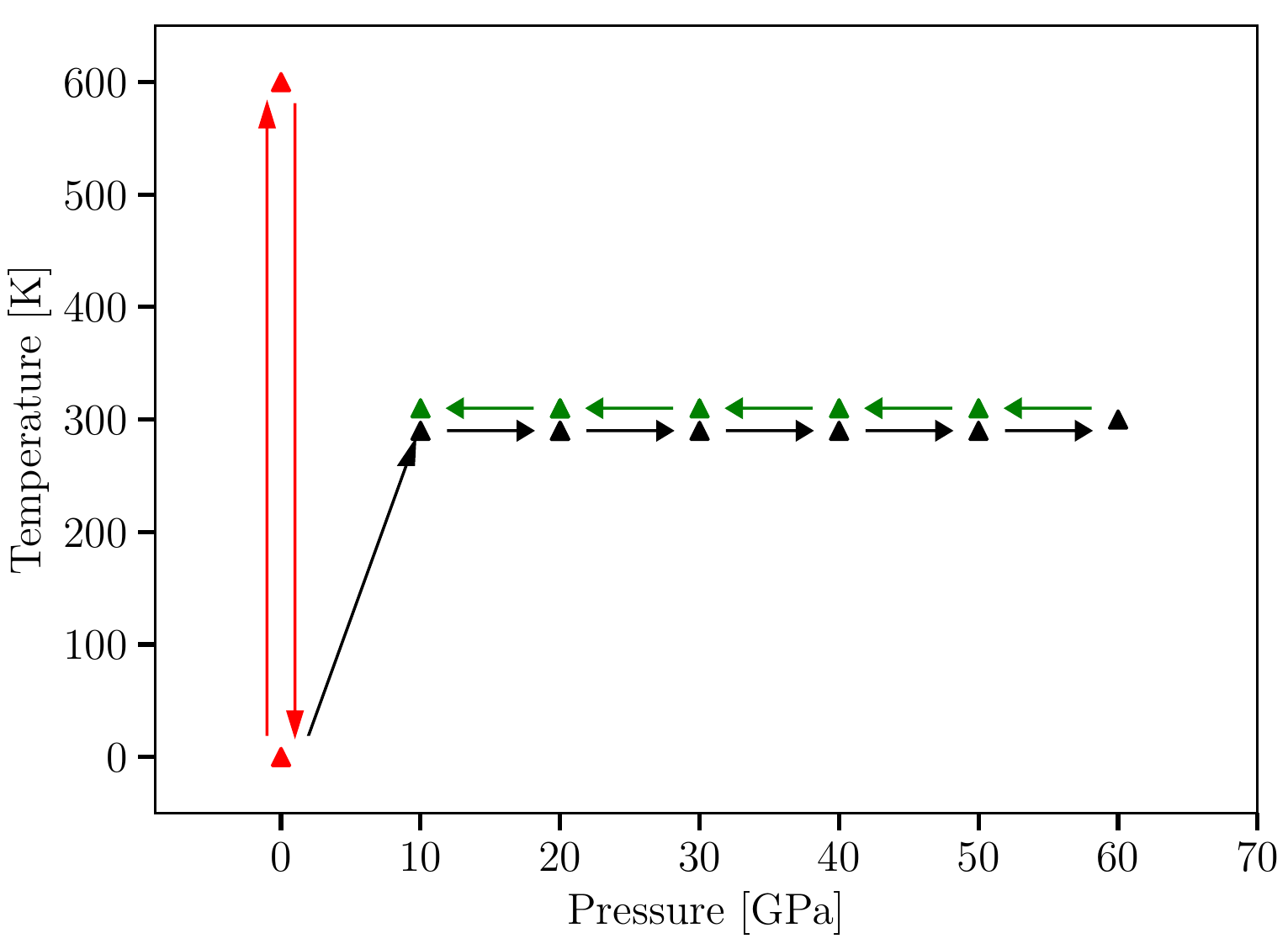}
\caption{Simulation protocol: red is preparation of amorphous SO\textsubscript{2} sample by heating perfect \textit{Aba} molecular crystal to 600 K and subsequent cooling to 0 K. Afterwards, the amorphous sample was compressed with 10 GPa steps (black) up to 60 GPa and then decompressed (green) with a reversed procedure.}
\label{fig:protocol}
\end{figure*}

\begin{figure*}%[tbhp]
\centering
\includegraphics[width=.95\linewidth]{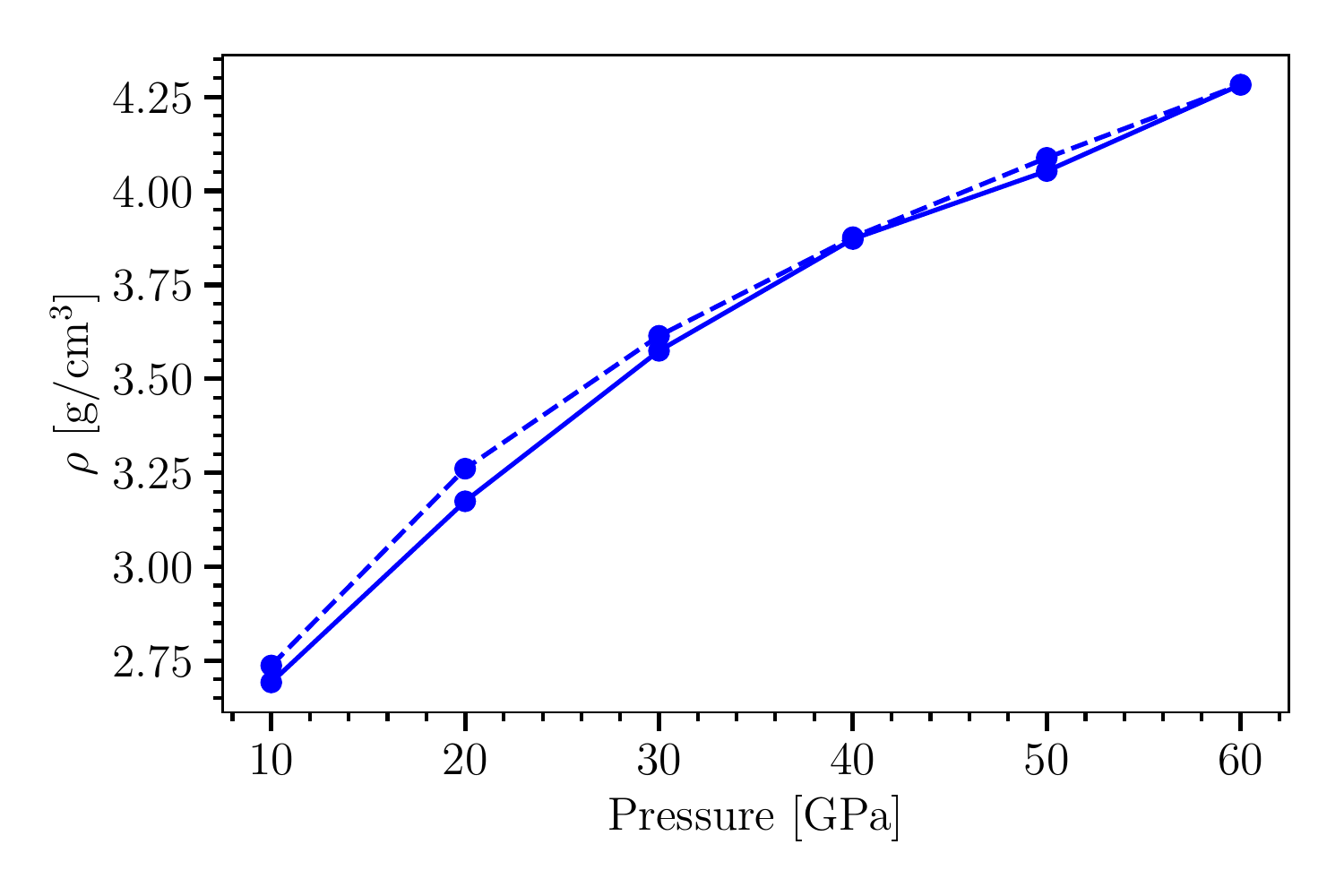}
\caption{Pressure dependence of density of SO\textsubscript{2} samples from simulations. Solid line represents density during compression and dashed line during decompression.}
\label{fig:density}
\end{figure*}

\begin{figure*}%[tbhp]
\centering
\includegraphics[width=.95\linewidth]{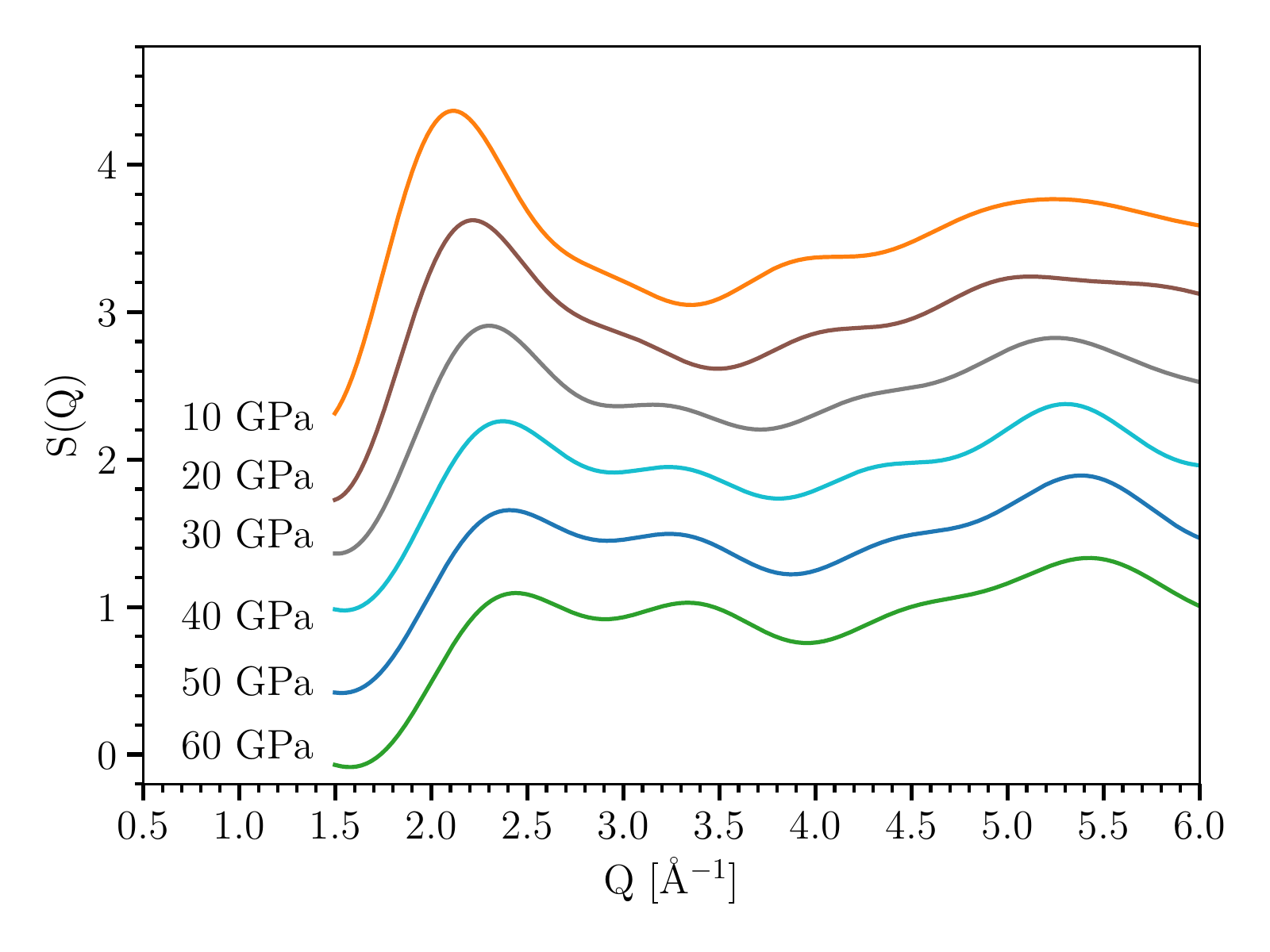}
\caption{Static structure factor of SO\textsubscript{2} samples during compression.}
\label{fig:sq_compression}
\end{figure*}

\begin{figure*}%[!htb]
	\centering
		\includegraphics[width=.49\linewidth]{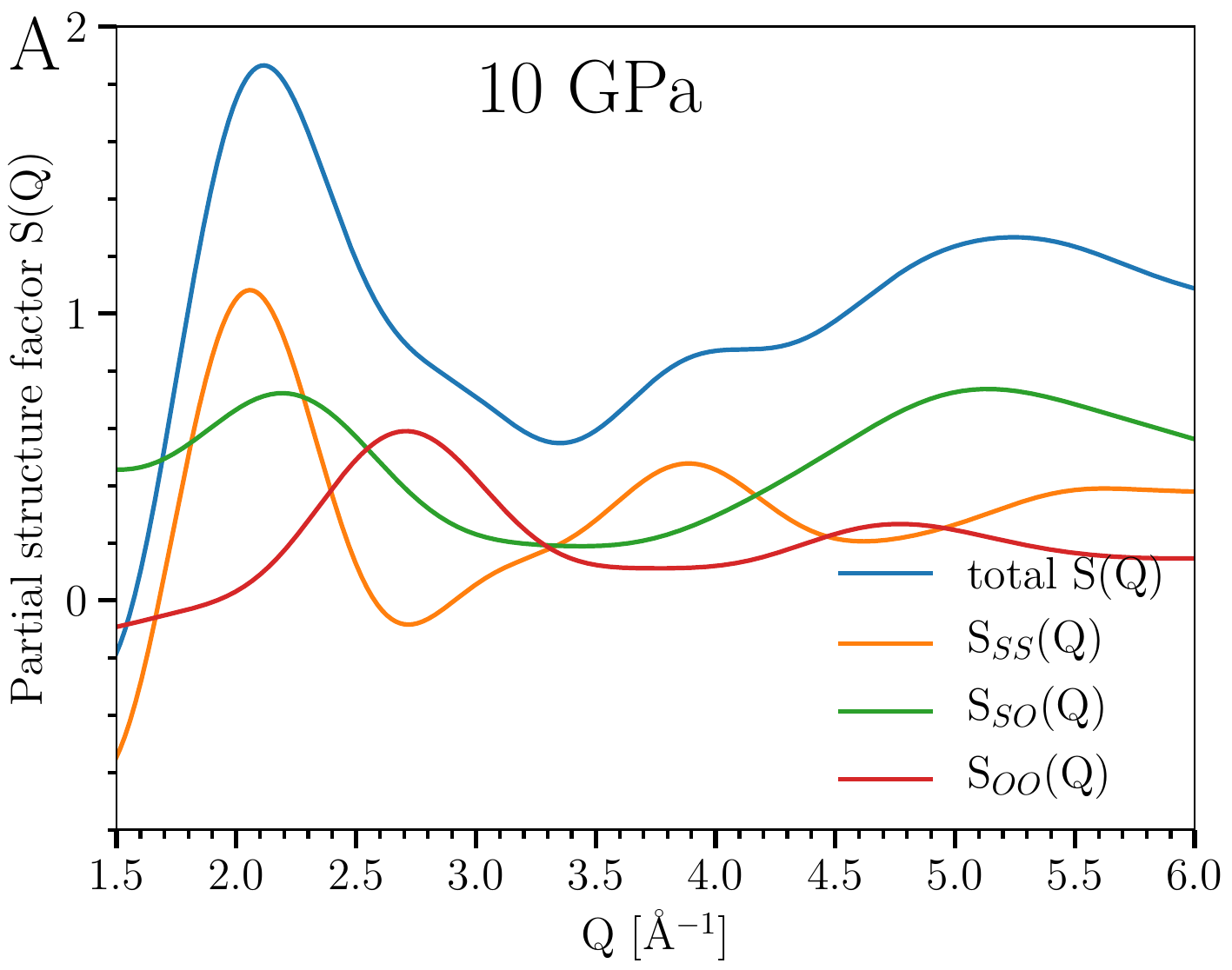}
		\includegraphics[width=.49\linewidth]{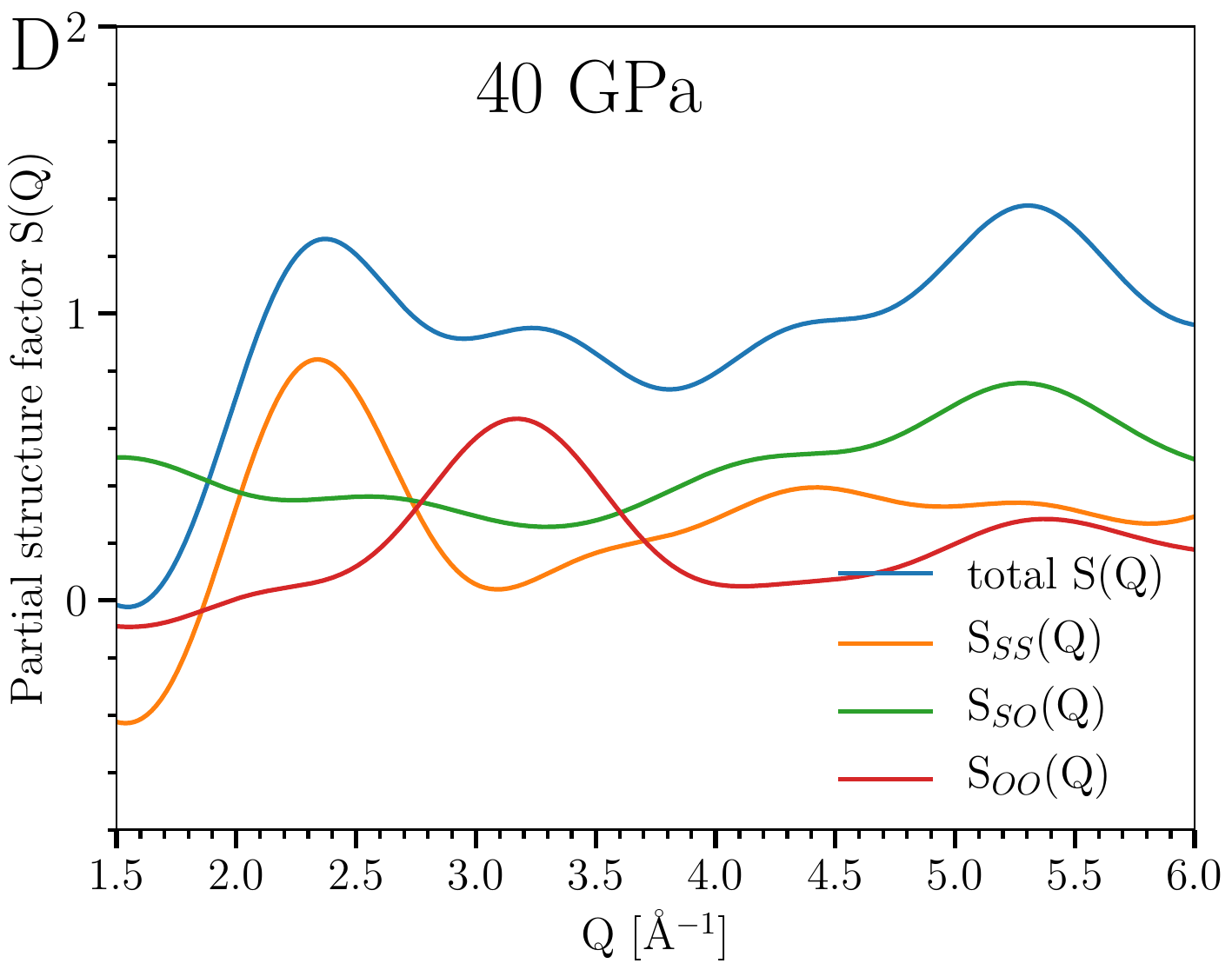}
		\includegraphics[width=.49\linewidth]{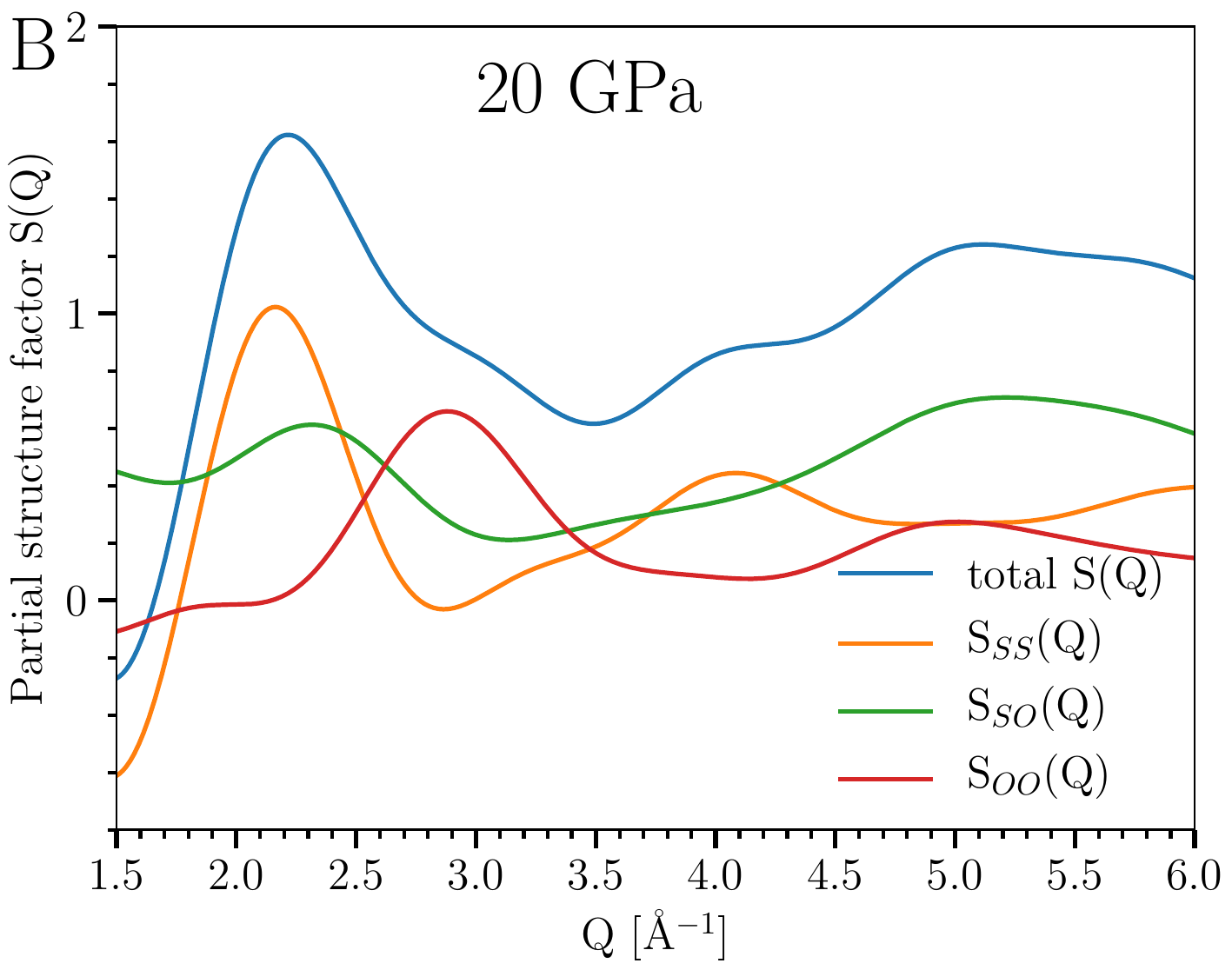}
		\includegraphics[width=.49\linewidth]{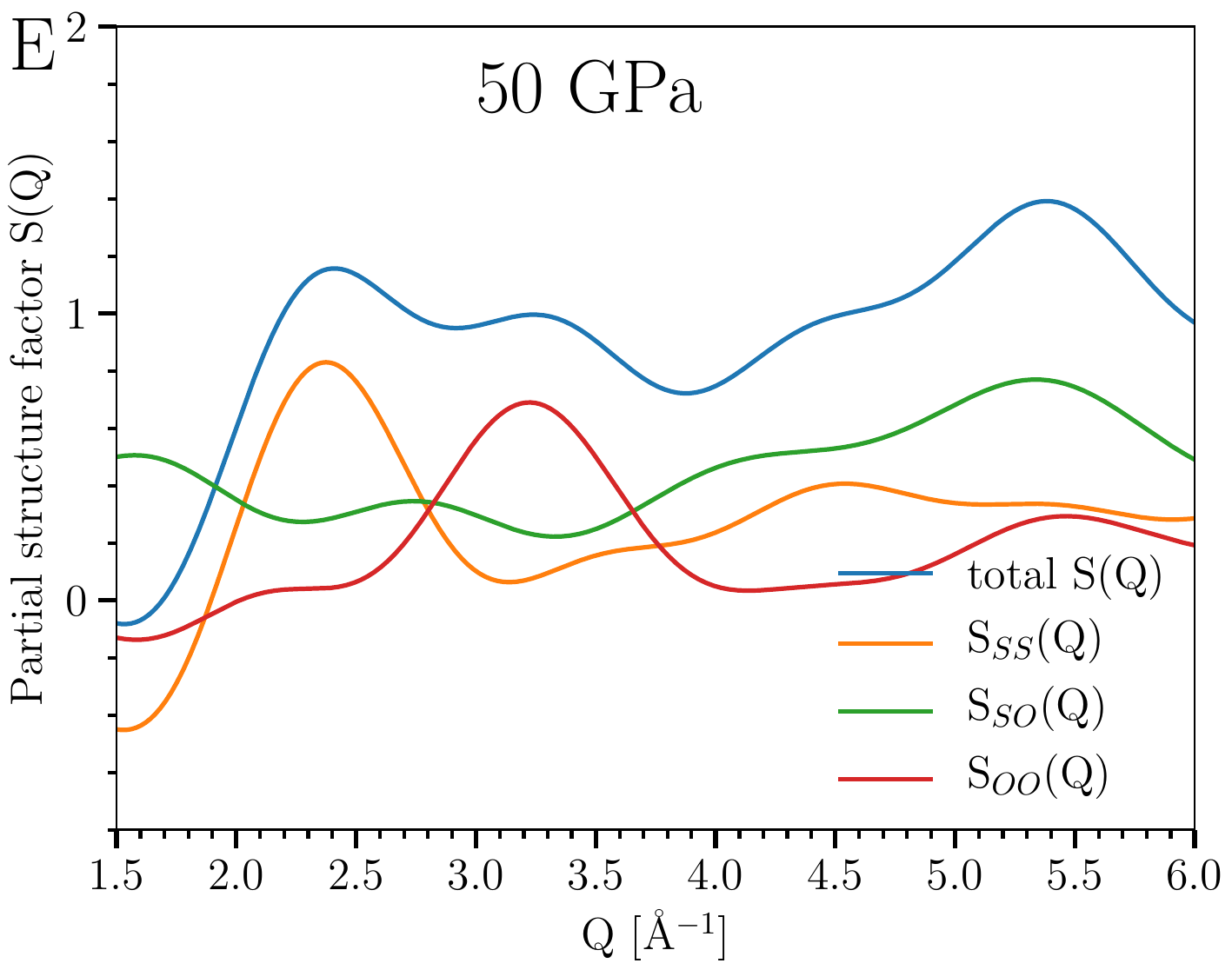}
		\includegraphics[width=.49\linewidth]{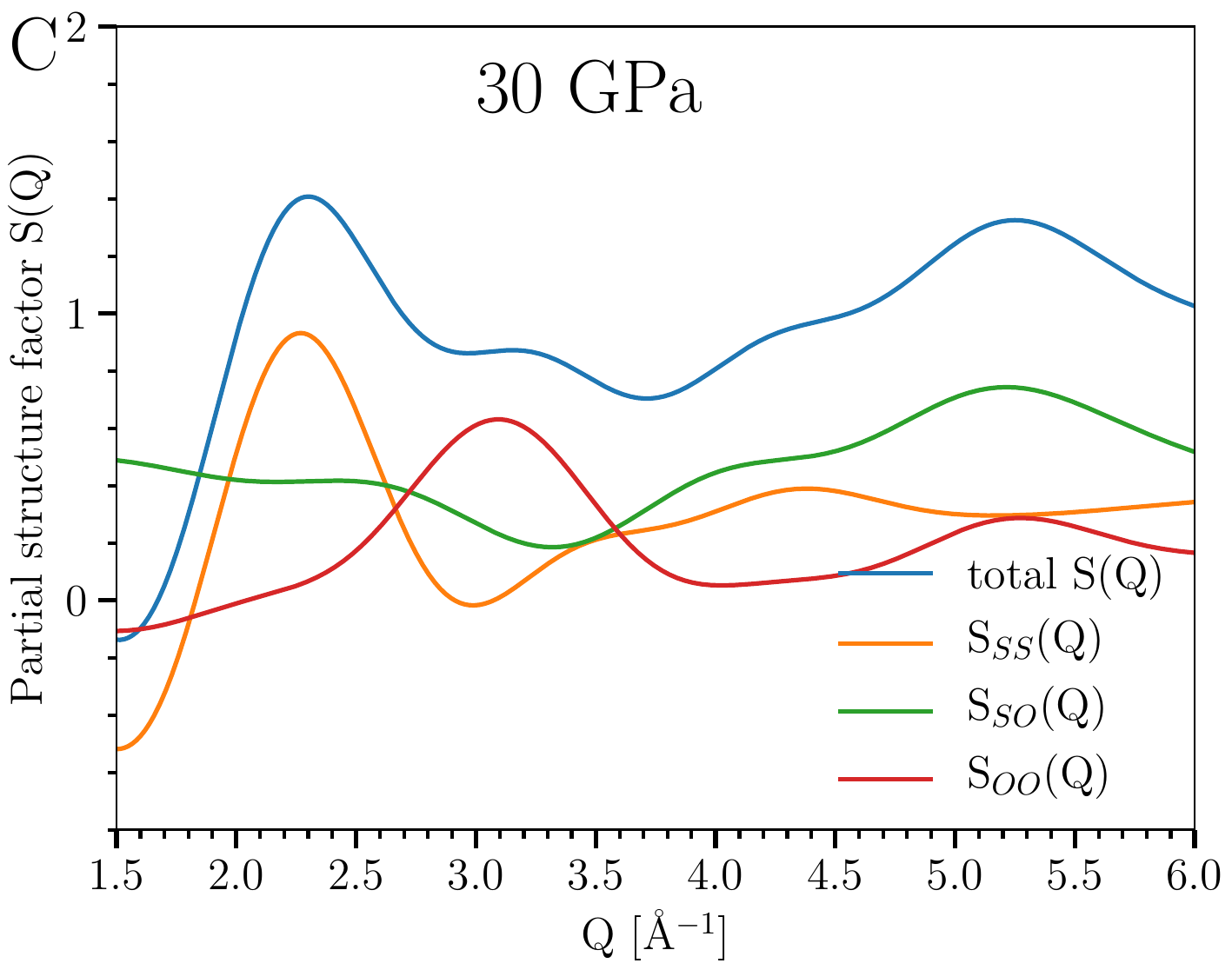}
		\includegraphics[width=.49\linewidth]{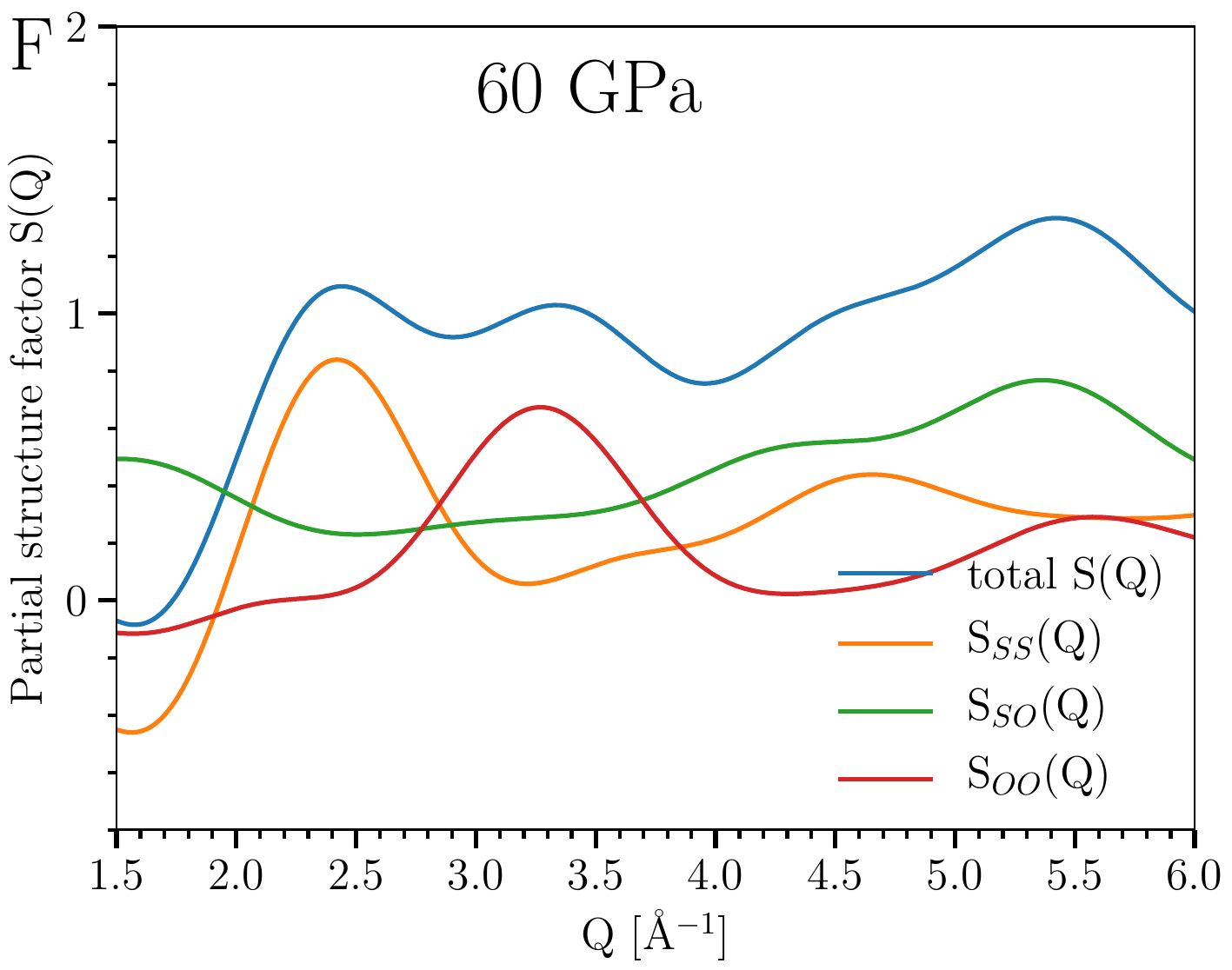}
	\caption{Partial structure factors $S_{\alpha\beta}(Q)$ from compression at pressures from 10 to 60 GPa with 10 GPa steps (A through F panels).}
	\label{fig:psf}
\end{figure*}

\begin{figure*}%[!htb]
	\centering
		\includegraphics[width=.49\linewidth]{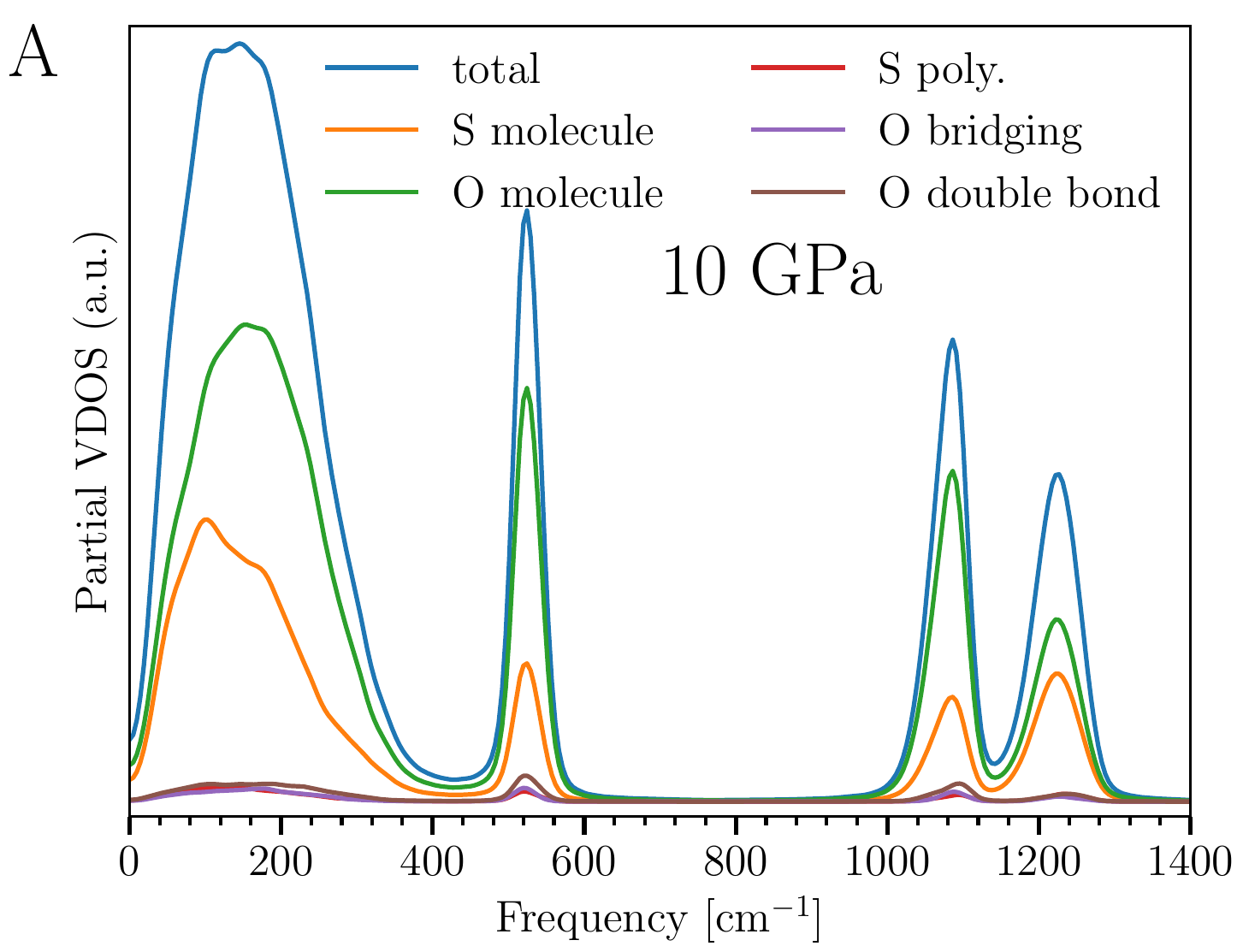}
		\includegraphics[width=.49\linewidth]{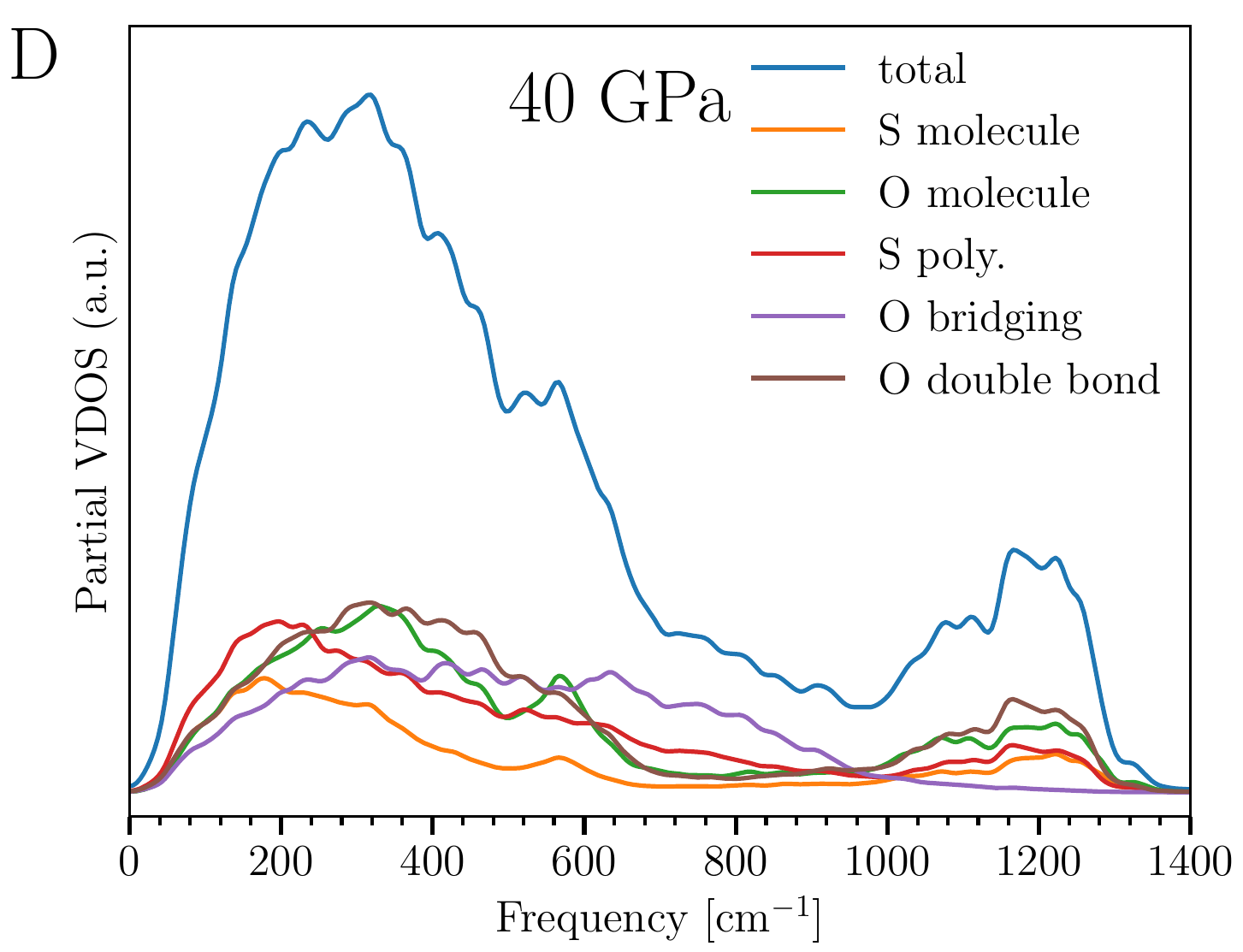}
		\includegraphics[width=.49\linewidth]{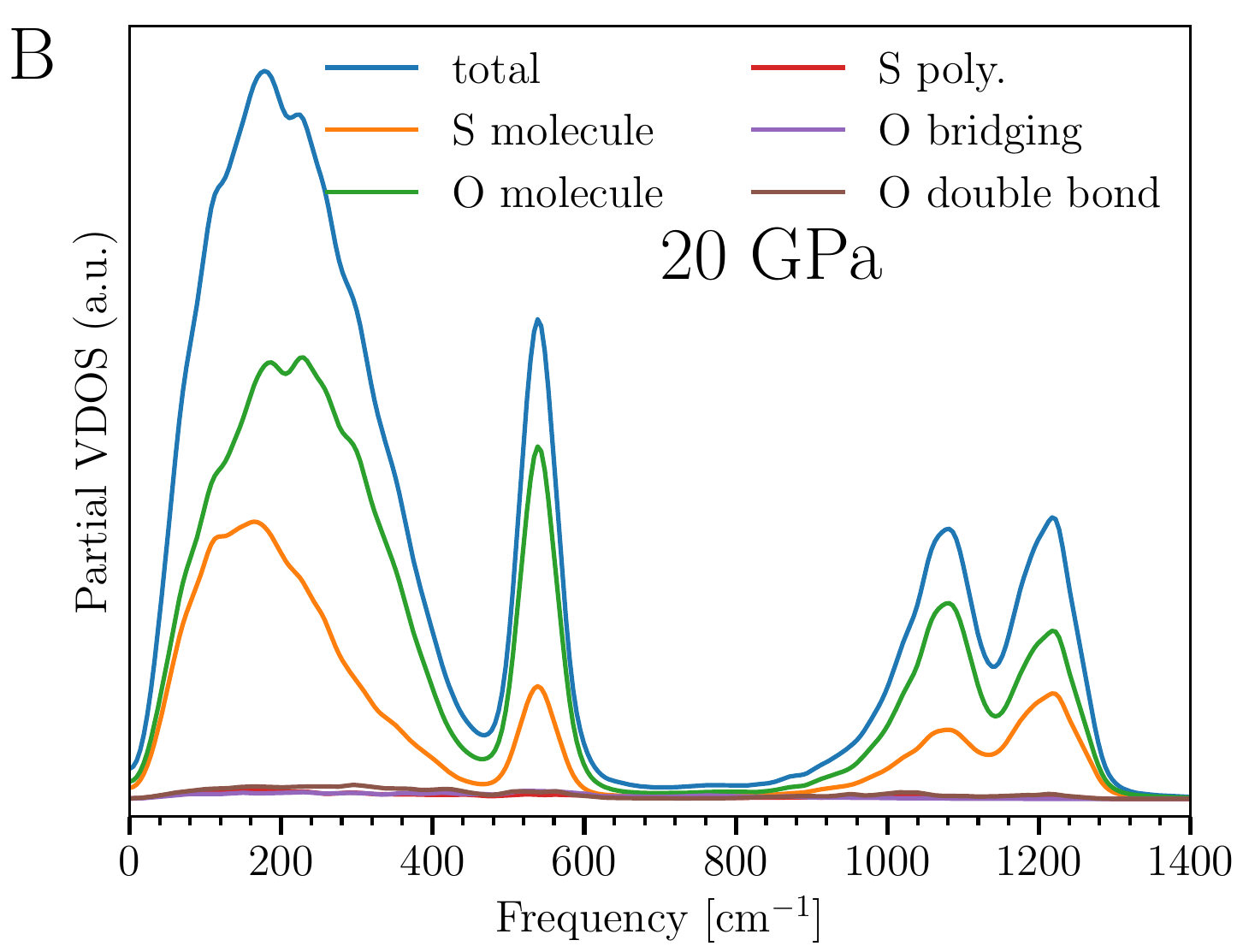}
		\includegraphics[width=.49\linewidth]{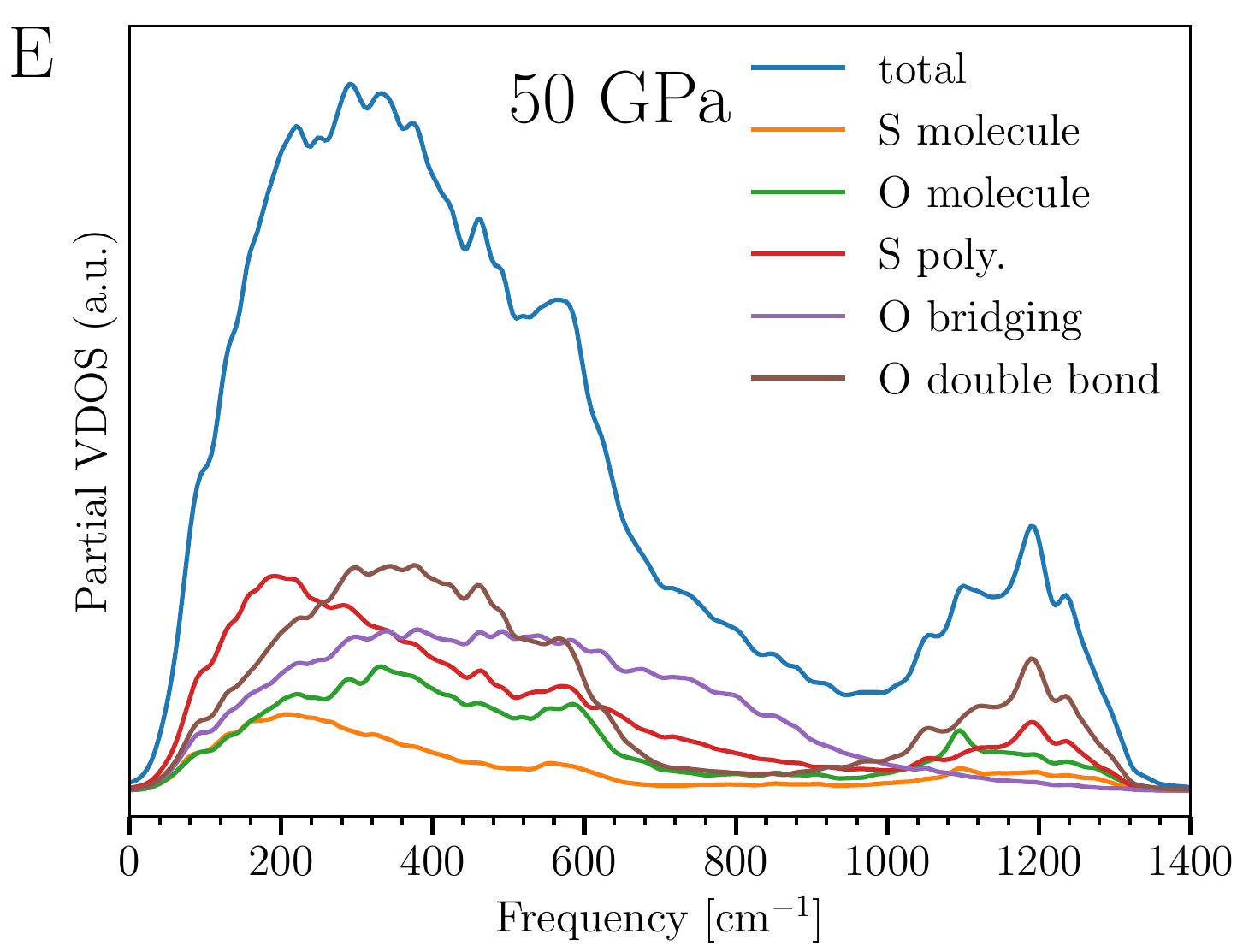}
		\includegraphics[width=.49\linewidth]{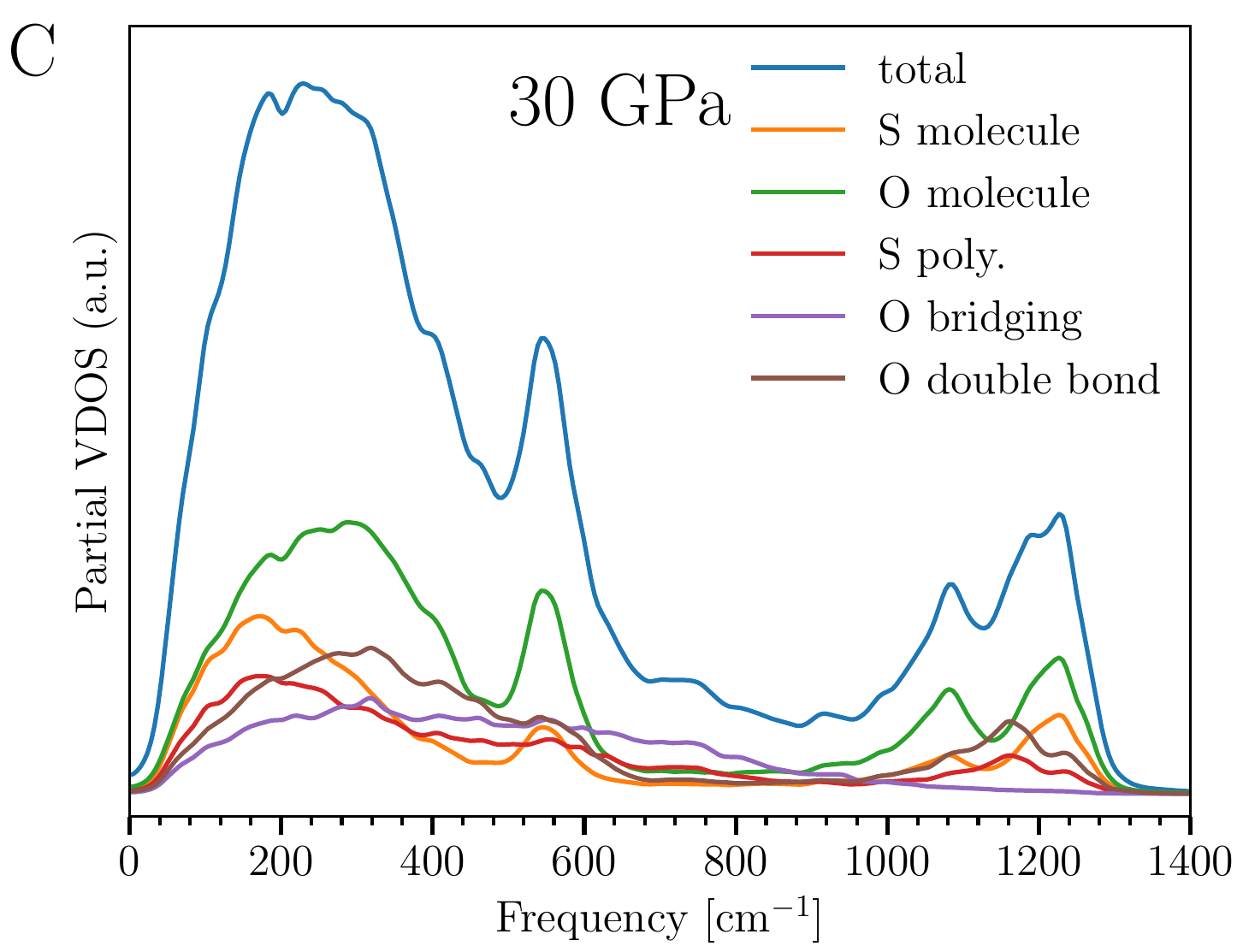}
		\includegraphics[width=.49\linewidth]{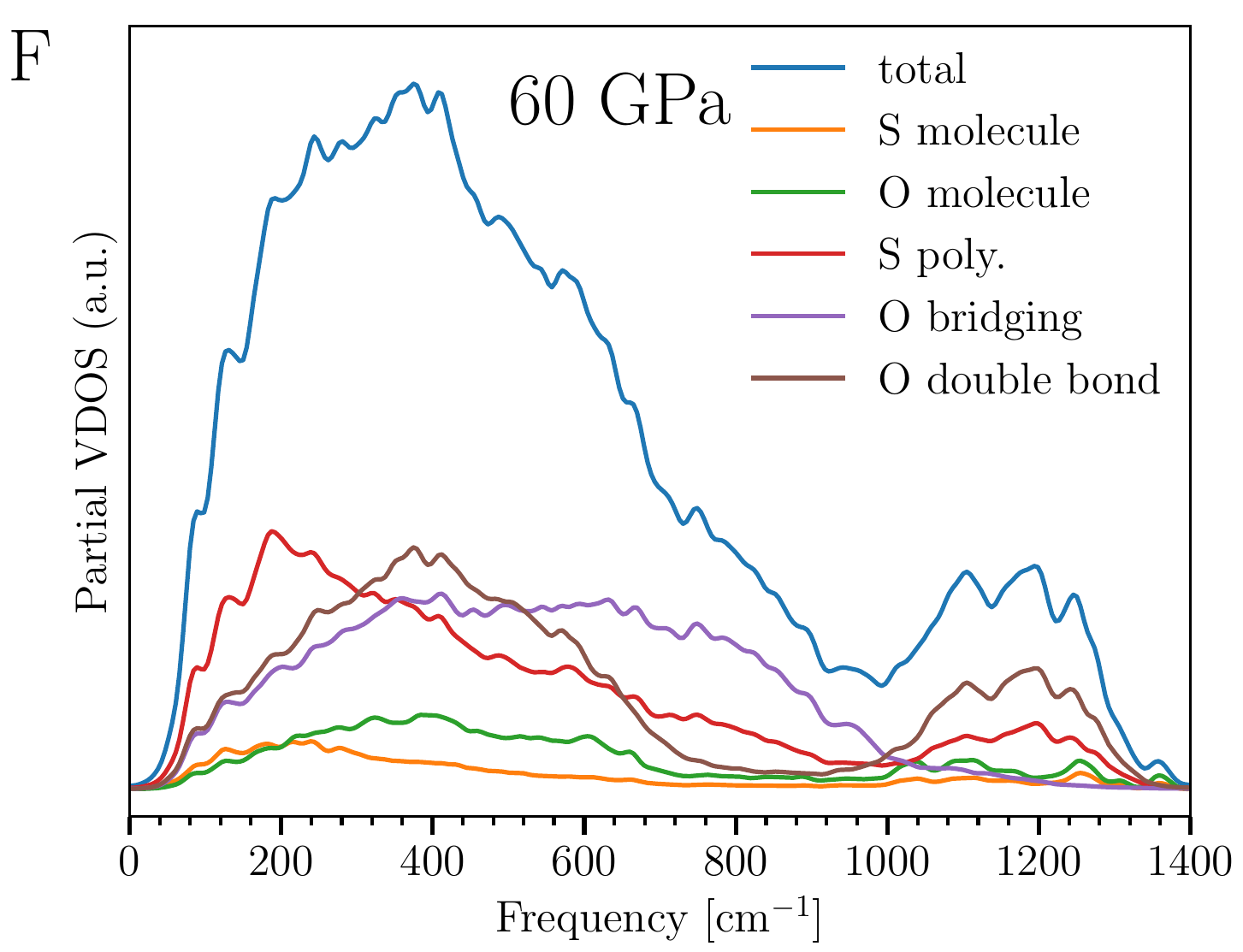}
	\caption{Partial VDOS from compression for pressures from 10 to 60 GPa with 10 GPa steps (A through F panels).}
	\label{fig:pVDOS}
\end{figure*}

\end{document}